\documentclass[twocolumn,reprint,aps,prd,nofootinbib,floatfix,superscriptaddress]{revtex4-1}
 \usepackage[unicode]{hyperref} 
\usepackage[utf8]{inputenc}
\usepackage{amsmath}
\usepackage{amsfonts}
\usepackage{amssymb}
\usepackage[left=2cm,right=2cm,top=2cm,bottom=2cm]{geometry}
\usepackage{braket}
\usepackage{dsfont}
\usepackage{hyperref}

\usepackage{graphicx}
\usepackage{tikz}
\usetikzlibrary{arrows,decorations.pathmorphing,backgrounds,positioning,fit,petri,shapes}
\usepackage{lipsum}

\usepackage{multirow}
\begin{document}

\title{Unsupervised Searches for Cosmological Parity Violation: Improving Detection Power with the Neural Field Scattering Transform}

\begin{abstract}
Recent studies using four-point correlations suggest a parity violation in the galaxy distribution, though the significance of these detections is sensitive to the choice of simulation used to model the noise properties of the galaxy distribution. In a recent paper, we introduce an unsupervised learning approach which offers an alternative method that avoids the dependence on mock catalogs, by learning parity violation directly from observational data. However, the Convolutional Neural Network (CNN) model utilized by our previous unsupervised approach struggles to extend to more realistic scenarios where data is limited. We propose a novel method, the Neural Field Scattering Transform (NFST), which enhances the Wavelet Scattering Transform (WST) technique by adding trainable filters, parameterized as a neural field. We first tune the NFST model to detect parity violation in a simplified dataset, then compare its performance against WST and CNN benchmarks across varied training set sizes. We find the NFST can detect parity violation with $4\times$ less data than the CNN and $32\times$ less than the WST. Furthermore, in cases with limited data the NFST can detect parity violation with up to $6\sigma$ confidence, where the WST and CNN fail to make any detection. We identify that the added flexibility of the NFST, and particularly the ability to learn asymmetric filters, as well as the specific symmetries built into the NFST architecture, contribute to its improved performance over the benchmark models. We further demonstrate that the NFST is readily interpretable, which is valuable for physical applications such as the detection of parity violation.

\end{abstract}

\author{Matthew Craigie}
\email{m.craigie@uq.edu.au}
\affiliation{School of Mathematics and Physics, The University of Queensland, QLD 4072, Australia}

\author{Peter L.~Taylor}

\affiliation{Center for Cosmology and AstroParticle Physics (CCAPP), The Ohio State University, Columbus, OH
43210, USA}
\affiliation{Department of Physics, The Ohio State University, Columbus, OH 43210, USA}
\affiliation{Department of Astronomy, The Ohio State University, Columbus, OH 43210, USA}

\author{Yuan-Sen Ting} 
\affiliation{Research School of Astronomy \& Astrophysics, Australian National University, Cotter Rd., Weston, ACT 2611, Australia}
\affiliation{School of Computing, Australian National University, Acton, ACT 2601, Australia}
\affiliation{Department of Astronomy, The Ohio State University, Columbus, OH 43210, USA}
\affiliation{Center for Cosmology and AstroParticle Physics (CCAPP), The Ohio State University, Columbus, OH
43210, USA}

\author{Carolina Cuesta-Lazaro}
\affiliation{Harvard-Smithsonian Center for Astrophysics, 60 Garden Street, Cambridge, MA 02138, USA}
\affiliation{The NSF AI Institute for Artificial Intelligence and Fundamental Interactions Massachusetts Institute of Technology, Cambridge, MA 02139, USA}
\affiliation{Department of Physics, Massachusetts Institute of Technology, Cambridge, MA 02139, USA}

\author{Rossana Ruggeri}
\affiliation{School of Mathematics and Physics, The University of Queensland, QLD 4072, Australia}

\author{Tamara M.\ Davis}
\affiliation{School of Mathematics and Physics, The University of Queensland, QLD 4072, Australia}

\maketitle


\section{Introduction}\label{sec:introduction}
The distribution of galaxies in the universe provides one of the most powerful probes in cosmology, encoding information from physics in both the early and late universe. The galaxy distribution offers a unique avenue to search for a violation of parity symmetry in the large-scale universe. Such a parity violation would manifest as a statistical discrepancy between the galaxy distribution and its parity-inverted counterpart.

In the early twentieth century, the conservation of parity symmetry was assumed to be ubiquitous in physics, but the discovery of parity violation in nuclear decay governed by the weak force \cite{wu1957} broadly called this assumption into question. Given that the weak force is not expected to play a role in the formation of the cosmological structure, the standard model of cosmology does not predict parity violation in the galaxy distribution. However, various extensions to the standard model could provide mechanisms that introduce parity violation into the galaxy distribution \citep{schmidt2015, jazayeri2023, cabass2023}. 

In 3D a parity violation is an asymmetry under an inversion of all three spatial coordinates. For the galaxy distribution, where each galaxy can be treated as a point in space, identifying such an asymmetry requires information from the 4-point correlations in the distribution. In 3D space, any structure consisting of three points (a triangle) lies on a plane. When spatial coordinates are inverted, the structure can be rotated out of the plane to return to its original configuration. In an isotropic field such as the galaxy distribution, triangle configurations are therefore indistinguishable from their mirror image. However, with four points or more, one can form a structure (a tetrahedron) that once mirrored, cannot be rotated back to its original configuration. Due to this property, \cite{cahn2022} suggested searching for parity violation in the parity-odd modes of the 4-point correlation function (4PCF). Subsequent applications of this technique to observational data from the Sloan Digital Sky Survey (SDSS) showed strong evidence of parity violation \citep{hou2023, philcox2022}. This approach is highly dependent on the simulated mock galaxy distribution catalogs that are used to model the noise properties of the 4PCF. Constructing these mocks accurately is challenging, given the associated computational difficulty of producing accurate N-body simulations, and the modelling difficulty of populating these simulations with galaxies in a way that is accurate and takes into account systematic effects that enter during observation. More recently, \cite{philcox2024} showed that the detection vanishes with the Uchuu-GLAM mocks \citep{ereza2023}, as opposed to the MultiDark-Patchy Mocks \citep{rodrigueztorres2016, kitaura2016} used in previous studies. This highlights the sensitivity of the 4PCF approach to the choice of mock catalogs, which poses a significant challenge for the detection of parity violation. 

As an alternative approach that avoids the dependence on mock catalogs, we frame the detection of parity violation in the galaxy distribution as an unsupervised learning task. This follows the approach developed for identifying parity violation in particle accelerators \citep{lester2022a, lester2022b, tombs2022}, and more recently our extension to cosmological parity violation \citep{taylor2023}. In this approach, we instead train a general machine learning model to identify the difference between the galaxy distribution and its parity-inverted mirror image. If the difference is also present in an unseen test subset of the data, then it is a property of the field that is not conserved under a parity symmetry. Provided we control for parity violation in the survey window function, and assume that the training and test volumes are uncorrelated, this difference constitutes a mock catalog-free detection of parity violation.

In the unsupervised approach, we require a general machine learning model that can extract the useful information from the galaxy distribution. In \cite{taylor2023}, we demonstrate a proof of concept by using a Convolutional Neural Network (CNN), which successfully identifies parity violation in a 2D toy model dataset. However, the CNN struggles when it is not provided large quantities of training data, which can be difficult to obtain in a more realistic cosmology context. This issue is exacerbated in three-dimensional space, where we find that the CNN is unable to detect parity violation for anything but the simplest data \citep{taylor2023}, even with large quantities of training data. This shortcoming motivates the exploration of alternative models that can identify parity violation in the galaxy distribution with less data and weaker signal-to-noise. 

The Wavelet Scattering Transform (WST) \citep{mallat2012} is a technique for summarizing information in a field that has recently grown in popularity for cosmological inference from the galaxy distribution \citep{valogiannis2022a, valogiannis2023, regaldosaintblancard2023, valogiannis2024}. The WST uses fixed wavelets as filters, and has explicit rotation and translation invariance built into its architecture, making it more robust to noise and suitable for capturing higher-order information in limited datasets. However, the standard WST's default filters are not ideal for extracting parity information. To better detect parity violation, we propose a novel method, the Neural Field Scattering Transform (NFST), that uses trainable filters parameterized with a neural field. This method captures the robustness of the WST architecture, while providing the filters with the flexibility to adapt to the task of parity violation detection.

In this paper, we demonstrate that the NFST is an effective tool for detecting parity violation in the galaxy distribution, with a focus on further developing the unsupervised learning method following the methodology and dataset described in \cite{taylor2023}. The paper is laid out as follows. In Section 2, we describe the WST and the novel NFST model. In Section 3, we introduce our dataset. In Section 4, we describe the unsupervised learning approach for detecting parity violation. In Section 5, we test variations of the NFST and benchmark it against a standard WST and CNN model. In Section 6, we discuss the implications and compare this approach with previous methods. In Section 7, we conclude and look forward.


\section{The Neural Field Scattering Transform}\label{sec:methods_nfst}
\subsection{The Wavelet Scattering Transform}
The Wavelet Scattering Transform (WST) is an operation developed by \cite{mallat2012} that compresses a field into a set of summary statistics. The WST uses iterative convolution operations with complex filters to extract high-order information from a field. It averages over symmetries and utilizes a stable complex magnitude non-linearity to ensure robustness to small distortions in the input, thereby enhancing its stability to noise.   

Given an input field, $\delta$, in configuration space, the WST first convolves $\delta$ with a complex wavelet filter, $\psi_a$, with specified scale and angle properties. This convolution produces a complex field that captures localized spatial features. Further taking a complex magnitude results in the first order scattering field, $I_a$. This field has high magnitudes in localized regions with structure that corresponds to the scale and angle of the filter, and low magnitudes elsewhere. The first-order scattering coefficient, $S^1_a$, is obtained by spatially averaging this field:
\begin{align}
    S^1_a = \langle{\left|\delta \star \psi_a \right|}\rangle = \langle I_a \rangle, 
\end{align}
where $\star$ denotes a convolution and $\langle \rangle$ denotes a spatial average. The spatial average over the first order scattering field captures how much structure in the field corresponds to the filter scale and angle, and reduces it to a single value. 

To extend the analysis to higher-order interactions, the WST iteratively applies additional convolutions. Convolving a first order scattering field $I_a$ with a second filter $\psi_b$ and taking the complex magnitude yields a second order scattering field $I_{ab}$. A spatial averaging then gives the second order scattering coefficient, 
\begin{align}
    S^2_{ab} = \langle{\left|\left|\delta \star \psi_a \right|\star \psi_b\right|}\rangle = \langle{\left|I_a\star \psi_b\right|}\rangle=\langle{I_{ab}}\rangle.
\end{align}
This second convolution captures the structure in $I_a$ that corresponds to the scale and angle of the second filter. As a result, $I_{ab}$ contains higher-order information, up to the 4-point correlation function. In principle, this process can be repeated to higher-order scattering coefficient, but for most physical fields, there is diminishing information gained above the second-order scattering coefficients \citep{cheng2021guide}. 

The ability of the WST to extract information relies on a carefully chosen set of filters that probe information in the field across a range of scales and angles. The filter set is typically built from a base filter $\Psi$. Individual filters in the set are dilations and rotations of the base filter, denoted $\psi_{jl}$, with subscripts $j$ and $l$ representing that filter's scale and direction indices, respectively. Each filter in the set is double the scale of the previous, corresponding to a dilation of $2^j$ compared to the base filter, up to a maximum scale $J$. This dyadic scaling introduces a logarithmic separation of frequencies that more evenly partitions high and low frequency information in natural fields. Each filter's direction is a rotation angle of $\theta_l=\pi l/L$ with respect to the base filter. In configuration space coordinates $\vec{r}$, the set of filters are related to the base filter through
\begin{equation}
    \psi_{jl}(\vec{r})=\Psi\left(2^j R_l \vec{r}\right),
\end{equation}
where $R_l$ is the rotation matrix for the angle corresponding to $l$. Alternatively, in Fourier coordinates $\vec{k}$, the set of filters are
\begin{equation}
    \hat{\psi}_{jl}(\vec{k})=\hat{\Psi}\left(2^{-j} R_l \vec{k}\right),
\end{equation} 
for the Fourier transform of the base filter $\hat{\Psi}$.

Using a set of these filters with the WST gives the first order coefficients for every combination of scale $j$ and angle $l$. Each filter $\psi_{jl}$ produces a first-order coefficient $S^1_{jl}$ which summarizes the amount of structure in the field corresponding to the filter's specific scale and angle. 

For the second order, there are scattering coefficients for each combination of $j_1, l_1, j_2, l_2$ where subscripts indicate the first and second filters, $\psi_{j_1 l_1}$ and $\psi_{j_2 l_2}$ respectively. The corresponding second-order coefficients $S^2_{j_1 l_1 j_2 l_2}$ summarize the structure that exhibits correlations between two different scales and angles in the field. In the second order scattering transform process, the first complex magnitude step pushes the field's information into scales larger than the first filter's scale. Applying a second filter smaller than the scale of the first filter therefore extracts no new information, so coefficients corresponding to $j_1 < j_2$ are not computed \citep{cheng2021guide}.

The WST coefficients form a compact summary of the clustering properties of the field, which is well suited for studying the galaxy distribution, which has complex coupling between scales and directions. Furthermore, the clustering summary probes the higher-order information in the field necessary to identify parity violation. However, the precise nature of the information extracted depends on the choice of the filter set. 

\subsection{Neural Field Filters}

\begin{figure}
    \centering
    \includegraphics[width=\columnwidth]{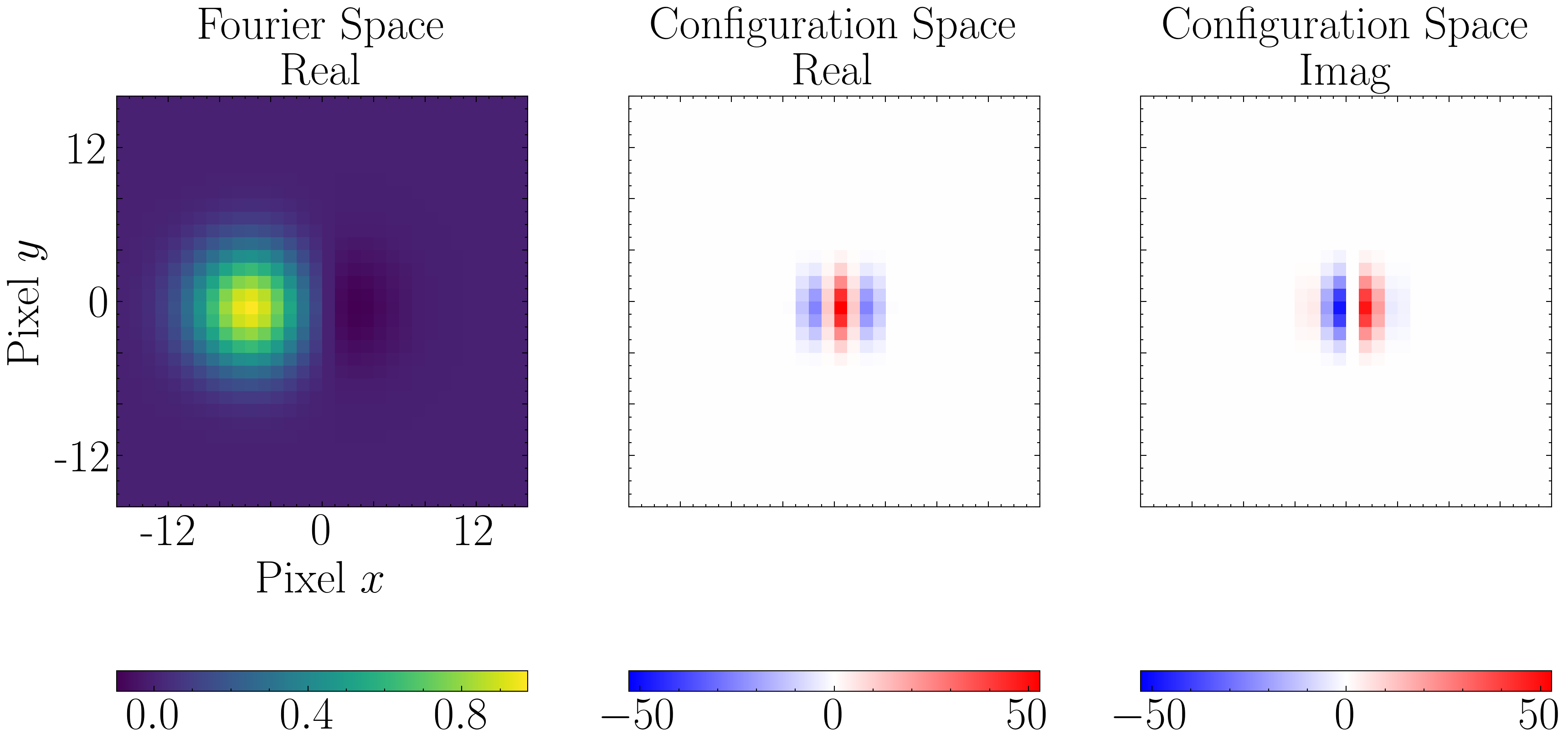}
    \caption{An example Morlet wavelet filter, which is the most common filter choice for the Wavelet Scattering Transform. The left panel shows the real component of the Morlet filter in Fourier space, in which the filter is a real-only, smooth directional bandpass. The middle and right panels show the Morlet filter in real and imaginary configuration space respectively, in which the filter is a 2D complex sinusoid modulated by a Gaussian envelope. This Morlet filter is placed on a $32\times32$ grid, and corresponds to scale parameter $j=1$ for this size.}
    \label{fig:morlet}
\end{figure}

The choice of the filter set $\psi_{jl}$ is crucial to the WST's performance. The most common choice of filters for the WST is a set of Morlet wavelets \citep{morlet1982}, an example of which is shown in Figure \ref{fig:morlet}. These filters are a set of complex wavelets with specific characteristic scales and directions corresponding to the $j$ and $l$ indices respectively. We describe the Morlet wavelets further in Appendix \ref{sec:appendix_morlet}. The wavelet structure maintains locality in both frequency and positional space which synergizes with the symmetries in the scattering transform architecture, which makes them well-suited to compressing a range of physical fields. However, this general nature means that they are not the optimal filter choice for specific tasks. 

We propose a modification of the WST such that a more optimized set of filters can be learned through gradient descent. Such an approach maintains the robust architecture of the WST, while providing some flexibility in the information extracted from the field. This approach can excel in cases like parity violation, where greater flexibility than the WST can offer is required, but there is not enough data available to train highly parameterized models like CNNs.

A pixel-wise parameterization, where each pixel in the filter is free to vary directly, is the typical choice for CNN models. For filters that are compatible with the WST however, we instead need a base filter that can be freely dilated and rotated. With a pixel-wise filter, this introduces resampling artifacts, leading to inconsistencies amongst the filter set. To overcome this problem, we instead parameterize the base filter as a neural field. Neural fields are neural networks that map an input positional coordinate to an output value, such that the field's information is held within the weights of the neural network \cite{xie2021}. They gained significant interest in graphics processing with the introduction of Neural Radiance Fields (NERFs) as a precise and compact neural field representation of a 3D scene \citep{mildenhall2020}. We encode the trainable filters as neural fields in Fourier space that take a frequency space coordinate as input and generate a real valued Fourier space filter value as output. Neural field filters can be trivially rotated and dilated by rotating and dilating the input Fourier space coordinates to construct the full filter set.  

\begin{figure}
    \centering
    \includegraphics[width=\columnwidth]{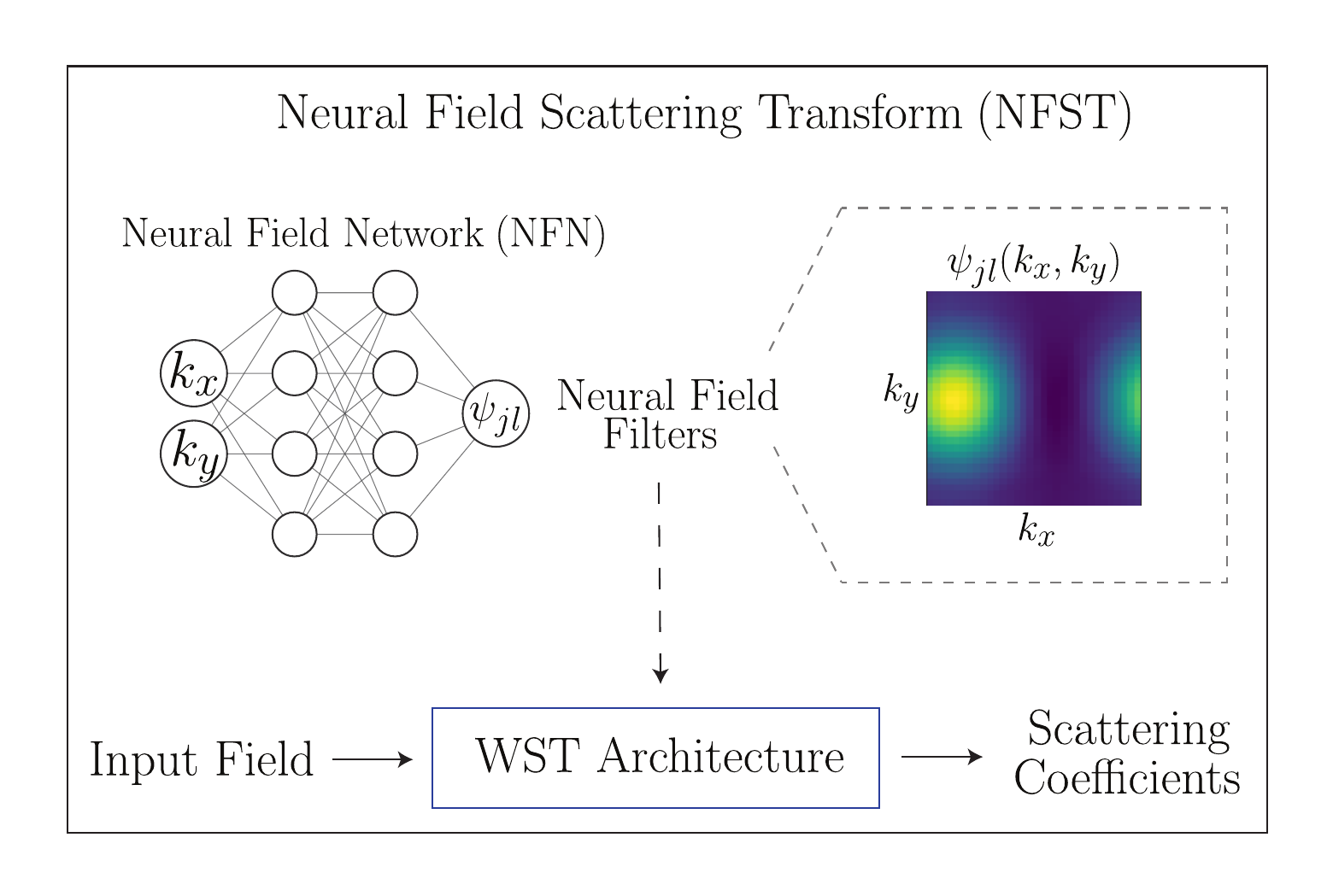}
    \caption{A schematic diagram of the Neural Field Scattering Transform (NFST) introduced in this paper, shown for a 2D case. A Neural Field Network (NFN) converts Fourier space input coordinates ($k_x, k_y$) into a set of Neural Field Filters ($\psi_{jl}$), dependent on the scale parameter $j$ and angle parameter $l$. The NFN carries the information about the filters in the weights of a simple fully connected neural network, which are learned throughout training. The NFST uses these filters within the architecture of the Wavelet Scattering Transform (WST), converting an input field into a compact and informative set of scattering coefficients in a way that can be optimized for a specific task like the detection of parity violation. In this example, the NFN is representing a Morlet filter. When trained on a specific task, the NFN can learn other structures which allows the NFST to better compress the information in the field.}
    \label{fig:nfst}
\end{figure}

Formally, the filters are parameterized as $\hat{\psi}_{jl}(\vec{k}) = F_{jl}(\vec{k})$, where the function $F_{jl}$ is a neural field built from a simple neural network that operates on a frequency coordinate $\vec{k}$, and returns a single scalar output. We choose to define the filters in Fourier space because they are real-valued, instead of the complex-valued configuration space filters. When discussing architectures, we refer to the neural network used to generate the neural field filters as the neural field network (NFN). We elaborate on the NFN formalism with Einstein summation notation in Appendix \ref{sec:appendix_nfn}. The resulting filters can be used directly in place of the Morlet wavelet filters in the usual WST architecture, which results in the technique we call the Neural Field Scattering Transform (NFST). We show a schematic of the NFST process in Figure \ref{fig:nfst}. We provide an implementation of the NFST as part of a publicly available general scattering transform package.\footnote{https://github.com/mattcraigie/scattering-transform}

There are a few further considerations when implementing the NFN within the NFST. We do not impose symmetry on the filters. Usually, the Morlet wavelets are symmetric under a rotation by $\pi$, so the filter set need only include rotations from $0$ to $\pi$ to capture, since anything beyond that will repeat the same information. Without this symmetry, the NFST filters must cover the full range from $0$ to $2\pi$, so we modify the rotation angle for the NFST to $\theta_l=2\pi l / L$. We additionally truncate the filters in Fourier space, such that for an input field size $N$, the $j$th filter is truncated so that frequencies above $N/2^j$ are cut. This ensures that each filter processes different scales of information, and ensures cross-scale interactions are probed. This mirrors the properties of the Morlet wavelet filters, for which the cut regions take the value of zero. This also provides a significant performance increase, avoiding the computation of Fourier space products that would always evaluate to zero.

\subsection{Parity-Sensitive Isotropic NFST}
The NFST summary of the field is a compression that retains directional information through the angular index $l$ for each coefficient. However, the galaxy distribution has no preferred direction. We can take advantage of this isotropy to improve signal-to-noise ratio by averaging the rotational symmetry.   

For first-order coefficients, this is straightforward. We directly average the $l$ indices, giving a single scattering coefficient for each scale $j$. Therefore, the coefficients represent the isotropic clustering in the field corresponding to the scale $j$.

Second order coefficients require more care because they contain two angular indices, $l_1$ and $l_2$. While it is common to average over all $l_1$ and $l_2$ for each $j_1$ and $j_2$ combination, this does not strictly perform an isotropic average. Instead, it mixes information where the angle between the first and second filters is different. As an alternative, we average over the angular difference $l_1 - l_2$, which does not mix different angles, similarly to the angular reduced WST described in \cite{allys2019}. We are careful not to destroy useful parity information by preserving the sign of the difference, which corresponds to a clockwise or anticlockwise angular difference between the two filters. 

This approach gives a parity-sensitive NFST that averages over the isotropy in the field while conserving all information. This compression improves the signal-to-noise, and enhances the performance of the NFST for isotropic fields.


\section{Parity Violating Toy Model Dataset}\label{sec:methods_data}
We test the NFST on a simplified parity violating dataset to demonstrate its potential, focusing on the 2D case. The general unsupervised learning approach detailed in Section \ref{sec:methods_unsup} extends to 3D without loss of generality. However, the CNN compression model component of the approach used in \cite{taylor2023} struggles with 3D data. For this reason, to build a comparison between the CNN and NFST models, we are limited to the 2D case in this work.\footnote{The NFST has been developed for its robustness, which helps it extend to the 3D case more effectively.}

\begin{figure}
    \centering
    \includegraphics[width=\columnwidth]{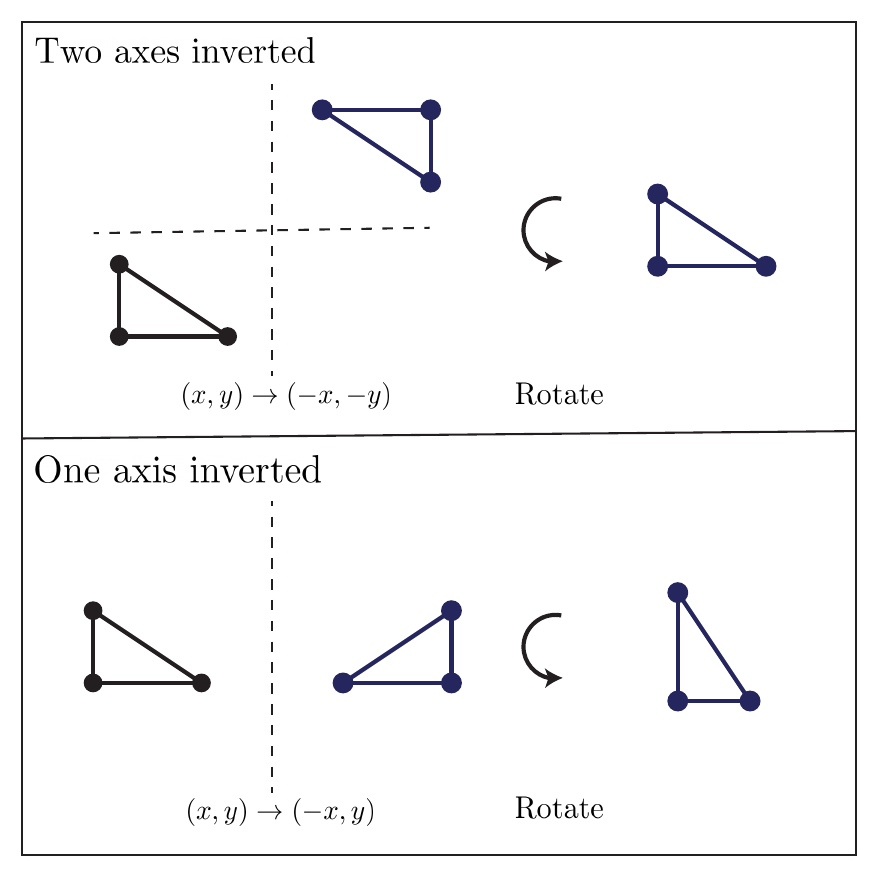}
    \caption{A demonstration of parity violating structures in two dimensions. After a coordinate inversion along both the $x$ and $y$ axes, a triangle can be rotated back into its original configuration (top panel). For a 2D isotropic field, the original and parity-inverted field will be statistically equivalent. However, a coordinate inversion along a single axis results in a triangle that cannot be rotated back to its original configuration (bottom panel). We use these triangles to add parity violation to the dataset.}
    \label{fig:triangles}
\end{figure}

This dataset is constructed following the approach from \cite{taylor2023} to approximate a homogeneous and isotropic galaxy distribution, where we manually imprint a parity violating signal in the dataset. In 2D, parity is not conserved under an inversion along a single coordinate. Inverting both axes is the same as a rotation by $180\deg$, so an isotropic field that has been inverted along both axes is statistically indistinguishable from its original configuration. In this 2D case, we only require 3-point structures to produce a parity violating field. Consider a right-angled triangle with two different arm lengths for the sides adjacent to the right-angle. Once mirrored along one axis, this triangle cannot be rotated back to its original configuration. Therefore, an isotropic field full of right angled triangles is distinguishable from its mirror, and is parity violating. We illustrate the parity violation of these triangles in Figure \ref{fig:triangles}.

Following this reasoning, we model the galaxy distribution by placing points in 2D space corresponding to the vertices of right-angled triangles. To create a parity violation, we only include right-angled triangles in the `right-handed' orientation and none in the `left-handed' mirrored orientation. When using the NFST, WST, and CNN models, we process data in smaller square portions of the full field, called patches, due to the models' limited input size. We generate these patches directly as $32\times 32$ pixel images. The image is generated by placing the points on a grid with a 2D histogram ranging from 0 to 32. 

\begin{figure}
    \centering
    \includegraphics[width=\columnwidth]{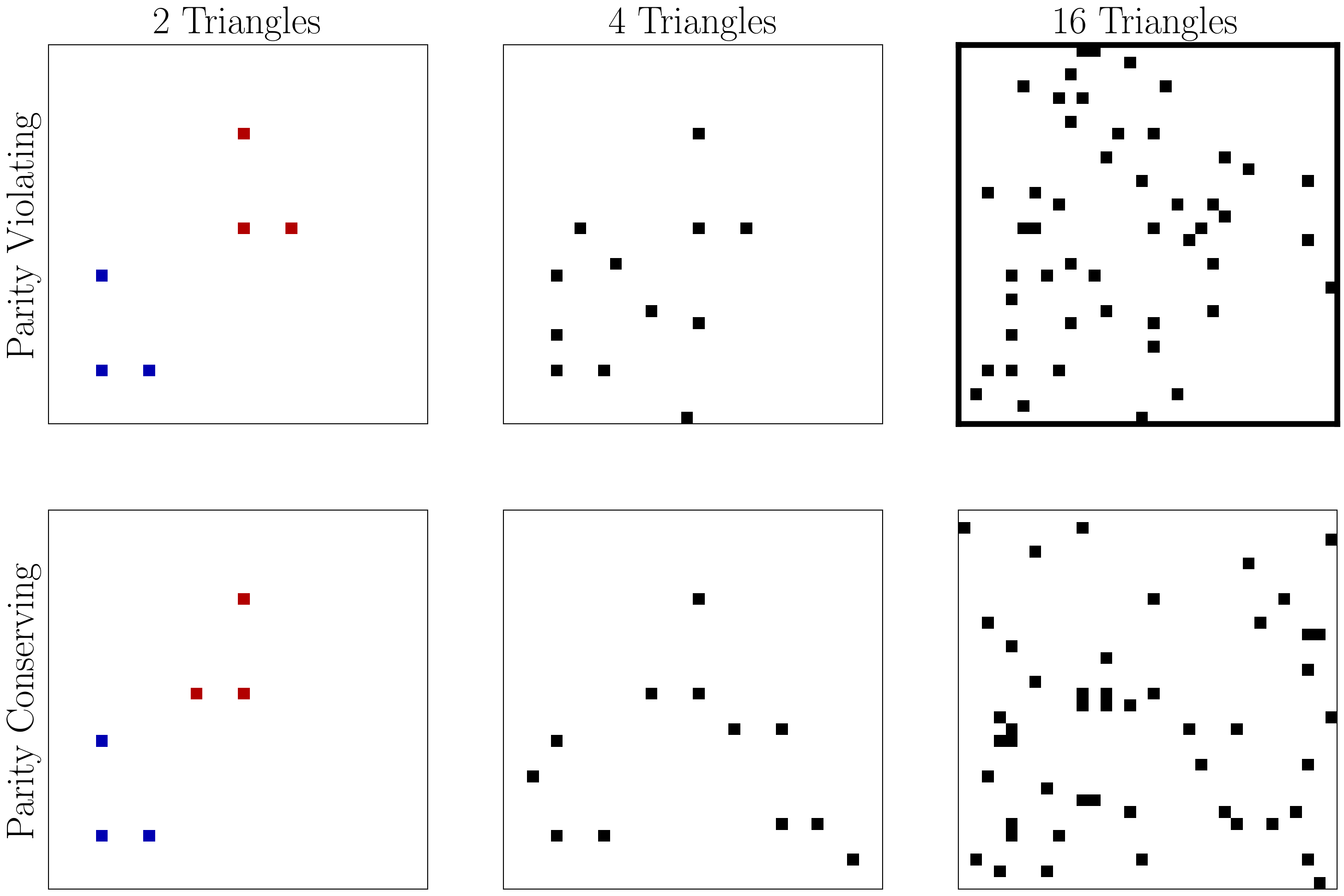}
    \caption{The top row shows the construction of an example patch from the parity violating dataset by placing right-handed triangles into a field. As a comparison, the bottom row shows a parity conserving patch constructed with equal left-handed and right-handed triangles. For 16 triangles, the parity violation is nearly impossible to discern by eye. The parity violating dataset in this work consists of patches matching the emboldened case of parity violating patches with 16 triangles.}
    \label{fig:mocks}
\end{figure}

For each patch, we place 16 triangles, keeping the distribution sparse with large empty gaps, to mimic a real galaxy distribution. The arm lengths of the triangles are 4 and 8 pixels, so a clear distinction can be made between the short and long sides. When placing each triangle the vertices may be outside the patch grid range, which would result in an inconsistent number of complete triangles and therefore differing parity violation signal in each patch. To account for this, we wrap any external points, ensuring that each patch has the same number of triangles. Note that this wrapping differs slightly in the treatment of boundary conditions compared to \cite{taylor2023}, where we instead deliberately placed triangles so that their edges would not extend beyond the boundary. The approach in that previous work results in a non-uniform density across the patch, so we choose to change it for the present work. Figure \ref{fig:mocks} shows the construction of a patch by placing the triangles at random. We also show a parity conserving patch for comparison, constructed by placing an equal number of left-handed and right-handed triangles. For fewer points, it may be possible to discern if the patches are parity violating by eye. However, the two patches are indistinguishable as we increase the number of points. 

\section{Unsupervised Learning to Detect Parity Violation}\label{sec:methods_unsup}

\subsection{Training an Unsupervised Model}
The overall goal of the unsupervised learning model is to compress all the parity violation information in an input patch into a single statistic. During training, the model learns to maximize the difference between the value of this statistic on the original and mirror image field.

For a scalar field $x(\vec{r})$ at a position $\vec{r}$, a parity violation is a discrepancy in the measurement of the field under a parity operation $P$. In the 2D case $P$ is a flip along a single axis, $x(r_x, r_y)\rightarrow x(-r_x, r_y)$ (or equivalently, $r_y \rightarrow -r_y$). Hereon, we drop the dependence on $\vec{r}$, writing $x(\vec{r})$ as $x$. Consider an arbitrary function of the field $g(x) \in \mathbb{R}$ that compresses the field into a single scalar statistic. We define $f(x)\in \mathbb{R}$ as the difference between $g(x)$ when applied to the field and its parity operated counterpart,
\begin{equation}\label{eq:parity_model}
    f(x) = g(x) - g(Px),
\end{equation}
If the field $x$ has a parity violation and the statistic $g(x)$ is sensitive to that parity violation, then $f(x)$ will be nonzero. This is the case we are interested in. 

\begin{figure}
    \centering
    \includegraphics[width=\columnwidth]{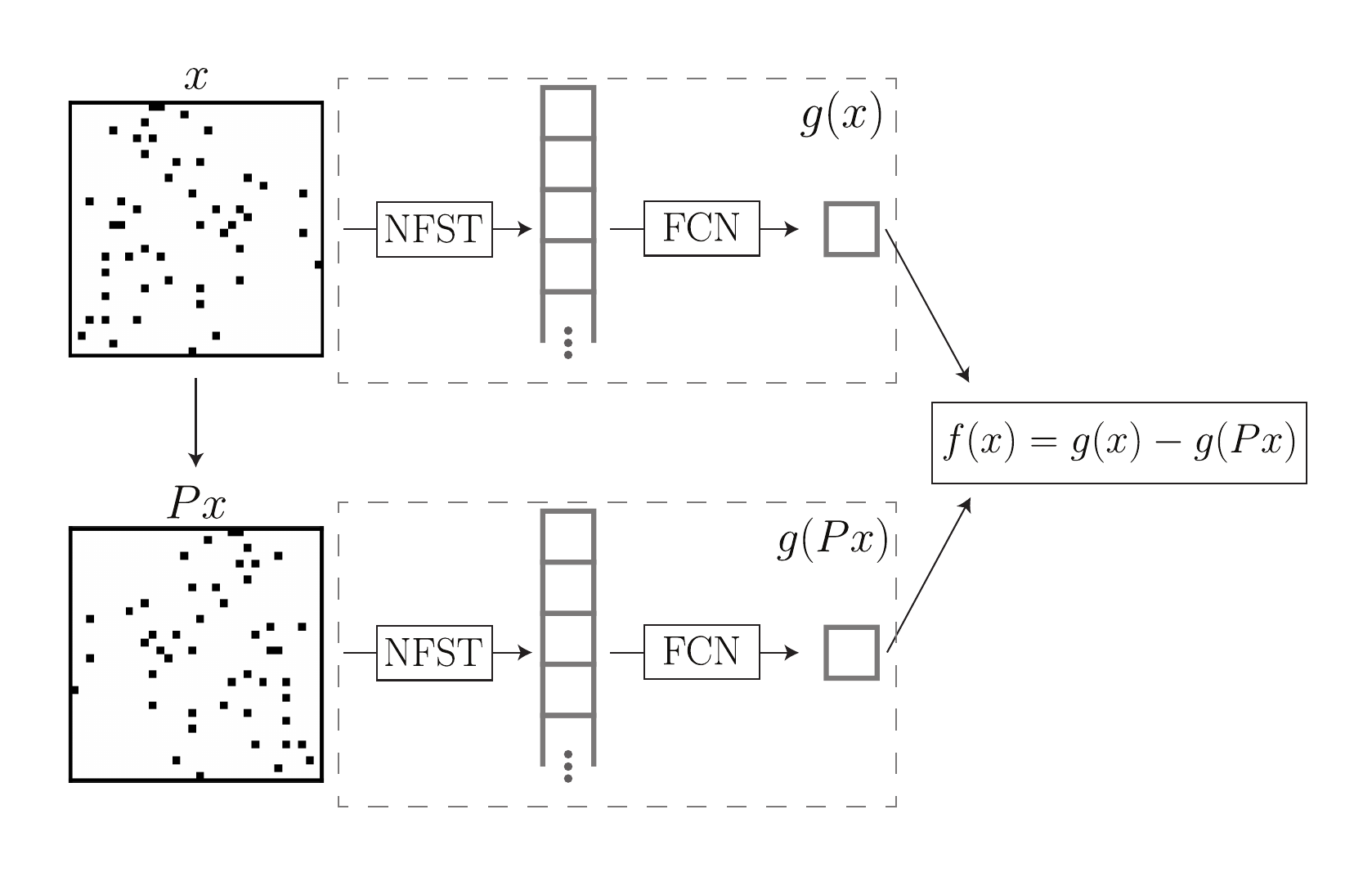}
    \caption{A schematic diagram of the unsupervised learning approach. An input field $x$ is compressed with the Neural Field Scattering Transofrm (NFST) to a set of scattering coefficients, and then compressed further with a Fully Connected Network (FCN) into a single output value. The full compression model is indicated by $g(\cdot)$. The process is repeated for the parity operated field $Px$, and the final parity violation statistic $f(x)$ is computed as the difference between these two values. In the unsupervised approach, the parameters of $g(\cdot)$ are learned throughout training, to maximize $f(x)$. We also test a Wavelet Scattering Transform (WST) model which replaces the NFST step with a WST, and a CNN model which serves as the entire $g(\cdot)$.}
    \label{fig:pv_schematic}
\end{figure}

Since the parity violation in $x$ could take any form, we use the unsupervised approach to learn $g(x)$ rather than trying to manually specify it. We let $g(x)$ be a trainable model and maximize $f(x)$ with gradient descent. The model will learn a new statistic $g(x)$ that maximizes the difference between the original field and the mirrored field. The resulting $g(x)$ is a parity summary statistic that is not conserved under a parity operation, and the resulting $f(x)$ is a statistic that represents the amount of parity violation in the field. A schematic of this operation applied to a single field is shown in Figure \ref{fig:pv_schematic}.

In practice, because the NFST, WST and CNN models work with smaller patches of the field, we use gradient descent to maximize $\mu_B$, the mean of $f(x)$ over the training batch of these patches. Since $f(x)$ can be trivially made arbitrarily large by scalar multiplication, we normalize by $\sigma_B$, the standard deviation of $f(x)$ values over the training batch. The loss function to be minimized is 
\begin{equation}\label{eq:loss}
    \mathcal{L}= - \frac{\mu_B}{\sigma_B}.
\end{equation}
With this loss function, minimizing the loss is equivalent to maximizing the signal to noise. 

Once the model for $g(x)$ has converged, we compute a parity statistic across the entire test set, denoted $\mu$, by taking the mean over the test set, 
\begin{equation}
    \mu = \langle f(x) \rangle_\text{Test}.
\end{equation}
A nonzero $\mu$ to statistical significance constitutes a detection of parity violation. More specifically, we wish to understand how confident we are that $\mu$ is nonzero due to the signal rather than natural variability that we find from taking a finite survey of cosmic structure, known as the cosmic variance. To approximate the cosmic variance, we use a bootstrapped distribution of $\mu$, computed by using resampled subsets of the test patches. This new bootstrapped distribution $\mu^*$ captures the variability of $\mu$ over the test set, which should approximate the variability of $\mu$ due to cosmic variance. From the bootstrapped distribution $\mu^*$, we compute the parity violation detection score $\eta$, 
\begin{equation}\label{eq:eta}
    \eta=\frac{\textrm{Mean}\left[\mu^*\right]}{\textrm{Std}\left[\mu^*\right]},
\end{equation}
which is the amount of standard deviations between the mean of the bootstrapped distribution and zero. This is effectively a $\sigma$-confidence level that $\mu$ is nonzero, so we consider $\eta > 3$ as a successful detection of parity violation. In this setup, $\eta$ is the $\sigma$-level at which we reject the null hypothesis that $\mu$ is consistent with zero. We elaborate on the validity of this approach for determining a confidence level in Appendix \ref{sec:appendix_verif}.

Equation \ref{eq:eta} can also be viewed as the square root of the $\chi^2$ for the case when the model expectation is zero, the data vector has been compressed into a scalar value, and the variance is measured not from external mocks. From this perspective, the model learns to estimate an analogue of the $\chi^2$ directly from the data, similarly to the $\chi^2$-based approach for detecting parity violation used in previous work \citep{cahn2022, hou2023, philcox2022, philcox2024}, but without the direct specification of the 4PCF.

\subsection{Training Setup}

\subsubsection{Full Compression Models}\label{sec:comp_models}

To examine the NFST's robustness and flexibility properties, we compare it to a benchmark standard WST model and a benchmark CNN model. We test the ability of these three models to detect parity violation in the simplified dataset described in section \ref{sec:methods_data}. 

For the NFST we choose $L=8$ angles, corresponding to 45 degrees between each filter rotation to balance between a fine coverage of angles without introducing extraneous coefficients. We choose $J=3$ scales, which means the largest Fourier filter is limited to a minimum wavelength of 8 pixels due to the Fourier space truncation. For larger $J$, the wavelength becomes comparable to the image size, and edge effects dominate over signal. After asymmetric angular averaging, this results in 49 coefficients.

We compare the NFST to an equivalent WST model with Morlet filters. For a fair comparison, we use $J=3$ and $L=4$ for the WST. This choice of $L$ is equivalent to $L=8$ for the NFST due to the rotational symmetry of the Morlet filter. We again use asymmetric angular averaging, which results in a total of 23 scattering coefficients.

The NFST and WST techniques output a compact set of summary statistics, but for the unsupervised approach, we must compress this further into a scalar value. For this compression, we use a simple fully-connected neural network consisting of a single hidden layer with 128 neurons. The hidden layer uses ReLU activation, and the output has no activation, resulting in a single scalar ouptut. The scattering transform models, followed by asymmetric angular averaging and then the compression network comprises the full parity violation detection model, which is $g(x)$ from equation \ref{eq:parity_model}. 

We also compare the NFST to a benchmark CNN model that follows the architecture described in \cite{taylor2023}, utilizing a deep convolutional network structure loosely based on AlexNet \citep{krizhevsky2012}. It utilizes 6 convolution layers and 2 max pooling layers, and ends in three fully connected layers with dropout layers used during training to improve robustness. While it is not necessarily the best CNN architecture possible, it was tuned in \cite{taylor2023} to detect parity violation for a very similar dataset so it provides a good benchmark to test the NFST against. The CNN outputs a scalar value, meaning the CNN serves as the entire compression model $g(x)$. We make a single change to the CNN by switching its padding mode to circular. This change is necessary because the NFST and WST both perform their convolution by using a product in Fourier space, which equates to a circular convolution in the spatial domain. Without this adjustment, the CNN could be at an unfair disadvantage compared to NFST and WST, since the patches are constructed with circular boundary conditions. 

\subsubsection{Training Strategy}

When applying the unsupervised approach to a real dataset, the data is limited in size. A large portion must be reserved for evaluating the detection significance, and must remain unseen during the training process -- this is the test set. In this proof of concept, we use a fixed test set size of 3000 mocks across all tests to maintain a fair comparison. 

For real data, once the test set is removed, the remaining data must then be partitioned into a training and validation set, where the number of samples in this combined set is given by $N_\textrm{TV}$. The training subset is used to directly train the model, while the validation subset is used to ensure the model is successfully capturing generalized patterns in the training subset and not over-fitting. Throughout, we use an 80\% training subset and 20\% validation subset split. 

We train the model by minimizing the loss over a training set with gradient descent. Throughout training, the validation loss is tracked and the model state with the lowest running validation loss is saved as a checkpoint. After training, we return the model to the state with the lowest validation loss and apply this to the unseen test set to evaluate the model's detection score $\eta$. We repeat the entire training process 10 times with a shuffled train and validation split and random initial conditions across trainable parameters. For each model, we take the maximum $\eta$ over these 10 runs as a fairer comparison, because an individual run may converge poorly by random chance. We use the maximum because in the practical setting with real data, we are free to repeat the training process multiple times and take the model with the best detection score. Even if a model generalizes poorly beyond the training set, we will still observe a $\eta<3$ over the test set, avoiding a false detection \citep{lester2022a}, and likewise, if $\eta>3$ over the test set we can be confident in a detection.

For training, we use an Adam optimizer with no scheduling. We tuned each model's training hyperparameters individually, but found the same setup was close to optimal in each case. For each model, we used 500 training epochs with a learning rate of $10^{-4}$ and a batch size of 64. The full implementation of the unsupervised learning approach is available publicly.\footnote{https://github.com/mattcraigie/UnsupervisedLearningPV} 


\section{Results}\label{sec:results}
\subsection{Tuning the Neural Field Filters}\label{sec:neural_field_tests}
\begin{figure}
    \centering
    \includegraphics[width=\columnwidth]{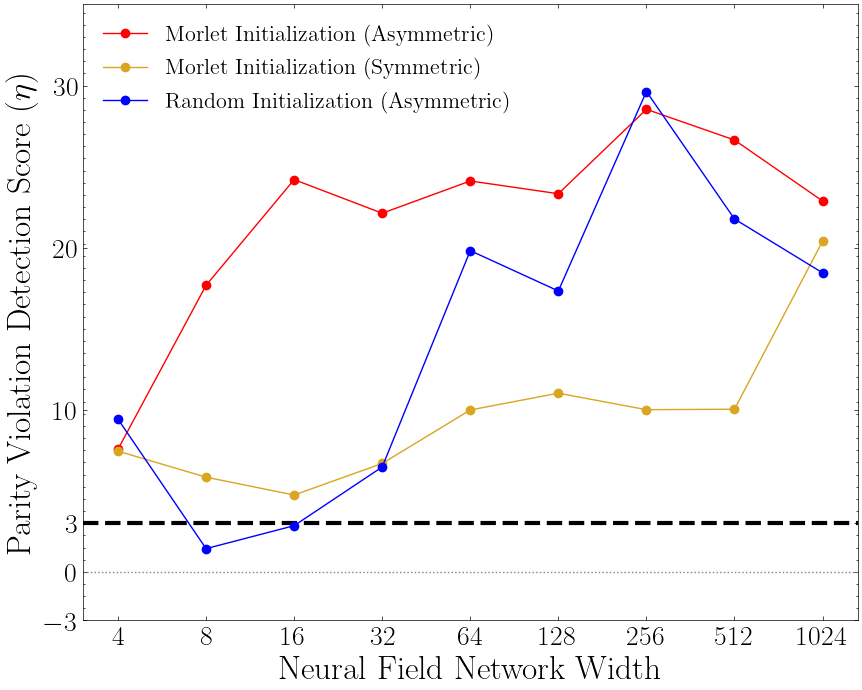}
    \caption{The parity violation detection score ($\eta$, Equation \ref{eq:eta}) for three Neural Field Scattering Transform (NFST) models across varied neural field network widths. The maximum $\eta$ across 10 tests with randomized initial conditions is indicated by the line. A higher $\eta$ indicates that the model is more effective at capturing parity violation in the simplified dataset. The model with Morlet initialization and asymmetry generally performs the best across different neural field network widths, and peaks in performance at a width of 256 neurons. The dashed black line at $\eta=3$ indicates the threshold at which we consider the model to have made a statistically significant detection.}
    \label{fig:nfst_sizes}
\end{figure}

We test variations on the NFN architecture to find the best setup for detecting parity violation in the dataset. The size of the NFST filters is fixed by the size of the input data, but we are free to vary the number of trainable parameters in the NFN. To test the effect of increasing the network size, we vary the width of the NFN between 4 and 1024 neurons per layer, testing all powers of 2 in between. In Figure \ref{fig:nfst_sizes} we show the performance of the NFST across a variety of NFN widths, with each width tested on a sample of 400 training and validation patches. For each test the NFN uses two fully-connected hidden layers. 

When randomly initialized, the NFNs produce neural fields that begin with no specific structure and cannot properly take advantage of the NFST's architecture. We test the effect of giving the filters a more suitable starting point by initializing the weights such that the NFN output matches that of a Morlet filter. To achieve this, we pre-train the NFN by minimizing the mean-squared difference between the NFN output and the base Morlet wavelet at each Fourier space coordinate using an Adam optimizer with a learning rate of $0.01$. We use a truncated version of the Morlet wavelet, which matches the truncated sizes of the NFST by wrapping the Morlet frequencies that extend beyond the truncation threshold periodically, a technique employed by WST packages like \texttt{kymatio} \citep{andreux2020kymatio}. We find a significant improvement with this Morlet initialization for most NFN widths, as shown in Figure \ref{fig:nfst_sizes}. With Morlet initialization, the model makes a strong $\eta>3$ detection for all NFN widths tested, from 8 to 1024 neurons. The model's performance peaks at 256 neurons, with $\eta=29$. We also find that the model converges better when initialized with a wider filter in Fourier space. As a comparison in Figure \ref{fig:nfst_sizes} we also show the performance when the NFN starts with a random initialization. Again, the performance peaks at 256 neurons with $\eta=30$, a marginal improvement. However, for all other NFN widths the random initialization performs worse than the Morlet initialization, excluding 4 neurons which also shows a marginal improvement. For this reason, we preference the Morlet initialization as the stronger, more consistent model. 

We also test a symmetric NFST model to explore the importance of filter asymmetry when detecting parity violation. In this model, we enforce a symmetric output along one axis in Fourier space by transforming the field coordinate inputs $k_x\to|k_x|$, before applying rotations. For a fairer comparison, we also apply Morlet initialization for this model. This model achieves a much lower detection score than its asymmetric counterpart at all NFN widths. It scores less than half that of its symmetric counterpart, except for NFN widths of 4 and 1024.


\subsection{Visualizing the Neural Field Filters}
\begin{figure}
    \centering
    \includegraphics[width=\columnwidth]{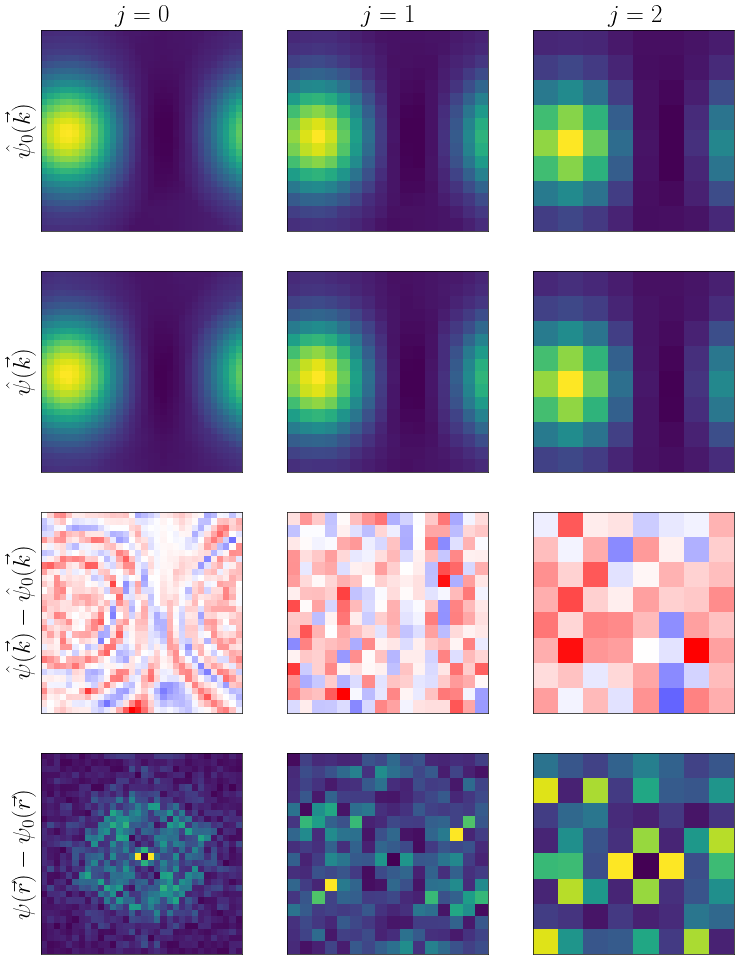}
    \caption{An example set of neural filters learned by the Neural Field Scattering Transform across the three scales of the filter indexed by $j$. Filters are truncated based on their scale, resulting in different sizes, and each filter is shown for a single orientation $l=0$. The top row is the initial state of the filter $\psi_0(\vec{k})$ after pre-training the neural field to match a truncated Morlet filter. The second row is the final state of the neural field $\psi(\vec{k})$ after learning to successfully detect parity violation. Visually, these are almost indistinguishable from the initialization. The third row shows the difference between final output and the state of the neural field after initialization $\psi_0(\vec{k})$, with positive differences shown in blue and negative differences shown in red. This emphasizes the learned structure, including prevalent concentric ring-like patterns. Note that the top three rows are in Fourier space and real-only. The bottom row is the absolute value of the difference of the inverse Fourier transforms of the final and initial states. Brighter regions indicate structures that are used to detect parity violation. The learned structure is similar to those that contribute to the parity violation in the dataset.}
    \label{fig:filters_k}
\end{figure}

Model interpretability is valuable for physical applications of deep learning. With the NFST, we can identify what features are being used for the NFST's decision, which is useful to confirm that the parity violation is indeed cosmological in origin and can aid in identifying the underlying physical cause. To visualize the features that contribute to the detection, in Figure \ref{fig:filters_k} we show the filters learned for the best model with Morlet initialization, which has an NFN width of 256 neurons. By eye, the filters appear to match the Morlet wavelet almost exactly. The average pixel-wise difference between the final learned filters and their Morlet initialization is only $\sim 2\%$ percent. 

To identify the non-Morlet structures, we show the difference between the initialized filters and the final filters. In these visualizations, frequencies that contribute most to parity violation are darker, with brighter red and brighter blue indicating regions where the filter is much higher or lower than the initialization respectively. There are some distinct learned features in these filters, most notably the concentric ring-like structures. There is also scale similarity between the filters, with similar ring-like structures appearing in the $j=0$ and $j=1$ filters. There are less clear structures in the $j=2$ filter, but high and low regions appear broadly in the same areas. The filters also learn clear asymmetries in Fourier space.

To relate these frequencies back to physical configurations in the field, we also show the magnitude of the inverse Fourier transforms of these frequency fields. Regions with higher magnitudes correlate to the structure that is relevant for parity violation detection.  For $j=0$, there is a significant structure in a band around 8 pixels away from the origin, with structure side lengths of 4 pixels visible. These scales correspond directly to the arm lengths of the parity violating triangles.  For $j=1$, there is a peak at 4 pixels from the origin in this truncated space, corresponding to the triangle arm length of 8 pixels in the input data space. For $j=2$, the peaks are less clear. Note that for each $j$, the zero-frequency mode has been set to zero for a clearer visualization.  

\subsection{Performance with Limited Training Data}

One of the NFST's primary goals is robustly detecting parity violation with limited training data. To test this, we compare the NFST's ability to detect parity violation in the dataset for a variety of training and validation set sizes $N_{\text{TV}}$ ranging between 50 and 3200 samples, with each sample size double the previous for a logarithmic spacing. For these tests, we use the best NFST model from Section \ref{sec:neural_field_tests}: a two-layer NFN, with 256 neurons in each layer and LeakyReLU activation, initialized with Morlet filters. We compare this NFST model with the standard WST model and a benchmark CNN model, as described in section \ref{sec:comp_models}. 

The NFST model makes a $\eta=3.8$ detection of parity violation for $N_{\text{TV}}=50$, and makes a very strong $\eta\gg3$ detection for all $N_{\text{TV}}\geq100$. The NFST's performance is strongly correlated with training set size. The WST model scores $\eta<3$ for all training set sizes $N_{\text{TV}}\leq800$, scoring marginal $\eta=3.1$ detections for $N_{\text{TV}}=1600$ and $N_{\text{TV}}=1600$. The WST shows some robustness in the limited data regime, scoring a borderline $\eta=2.9$ for $N_{\text{TV}}=100$. In general, the CNN performs better, with $\eta>3$ detections for $N_{\text{TV}}\geq200$. Like the NFST, the CNN performance scales with increasing data, but with significantly lower score at each $N_{\text{TV}}$.

The detection threshold of each model is the $N_{\text{TV}}$ required for the model to make a detection, which is visually equivalent to where the model's line crosses the black dashed line representing a $3\sigma$ significance of detection. For the NFST, this occurs at the lowest training set size tested, $N_{\text{TV}}=50$. For the CNN model, this is approximately $N_{\text{TV}}=200$, or 4 times the training data. For the WST, this is approximately $N_{\text{TV}}=1600$, or 32 times the training data. However, the WST's near detection occurred with $N_{\text{TV}}=100$, or $2\times$ the training data.  

We emphasize the result that the NFST can make a strong detection in cases where the WST and CNN cannot. This occurs at $N_{\text{TV}}=50$ and $N_{\text{TV}}=100$, where the NFST scores $\eta=3.8$ and $\eta=10$ respectively, while the WST and CNN score $\eta<3$. In the context of real data, this would mean the WST and CNN models missed the parity violating information in the field, while the NFST captured it.

\begin{figure}
    \centering
    \includegraphics[width=\columnwidth]{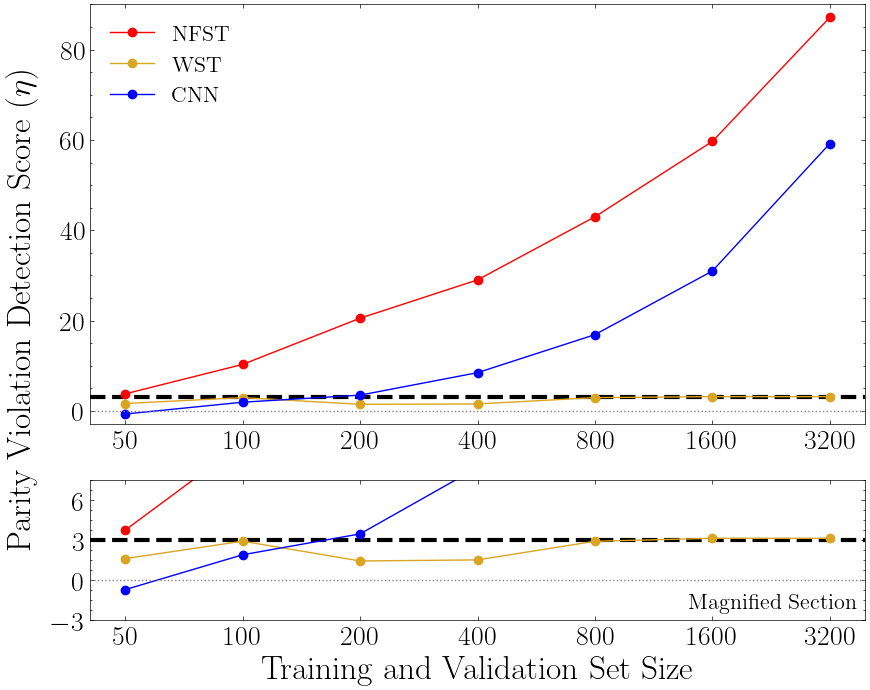}
    \caption{The parity violation detection score ($\eta$) for three different detection models for various amounts of training and validation data. The maximum $\eta$ across 10 tests with randomized initial conditions is indicated by the line. A higher $\eta$ indicates that the model is more effective at capturing parity violation. The Neural Field Scattering Transform (NFST) is the best across all training and validation set sizes, outperforming the standard Wavelet Scattering Transform (WST) and a Convolutional Neural Network (CNN) benchmark model. As in Figure \ref{fig:nfst_sizes}, the dashed black line at $\eta=3$ indicates the threshold at which we consider the model to have made a statistically significant detection. The lower plot shows a magnification of the lower section of the first plot.}
    \label{fig:datascaling}
\end{figure}


\section{Discussion}\label{sec:discussion}
\subsection{Unsupervised Learning}

Previous work has explored two approaches to parity violation detection with the 4PCF, namely a $\chi^2$ likelihood test and a rank test, both of which depend on simulations to accurately capture the noise properties of the galaxy distribution. The main challenge with these approaches is the computational difficulty in producing a large suite of simulations that are accurate enough to model the dark matter halo distribution down to small scales. Beyond the computational challenges there are other difficulties, including uncertainties in the theoretical modelling of the halo-galaxy connection, and difficulties associated with injecting systematics present in realistic data into a forward-modelled mock catalog. The core advantage of the unsupervised learning approach is the ability to compress all the information in the field into a single value, including higher order information from beyond the 4PCF. Ideally, this compact and information-rich representation allows the variability to be sufficiently well-estimated from the data alone. 

For this approach to work with data, we assume that the patches of data are independent. If subvolumes are correlated through long wavelength modes larger than the patch size, then the variance of $\mu$ will have an additional covariance term between subvolumes. As a result, the detection score $\eta$, which depends on the variance of $\mu$ across the test dataset, will be incorrectly specified, which could lead to a false detection. We can mitigate these effects by sampling the train and test sets from different regions in the survey, but before applications to data, we will need to verify that this holds with realistic parity violating mock universes in 3D, beyond the simplified 2D data used in this work. 

If this does not hold, we can also estimate the variability using simulations to understand correlations between the large-scale and small-scale modes. Since we only need to estimate the variance of a single parameter instead of the full covariance of the 4PCF modes, this will require orders of magnitude fewer simulations than previous methods. For the same compute time, this allows more accurate simulations, and an overall more accurate measurement of the parity violation. 

Another difficulty of the unsupervised approach, compared to the previous approaches, is the handling of the survey window. The observed distribution will naturally have a parity violating window function, which will dominate any detection model. In future data applications, we anticipate the best way to solve this problem is to cut the real data to a parity symmetric window. This will sacrifice some data volume, but the advantages gained by the unsupervised approach's escape from simulations should outweigh this disadvantage.

\subsection{Inductive Biases}
In broader machine learning and deep learning, there are two common trends to improve model generalizability. The first is to scale up the number of trainable parameters of the model, which has been shown to lead to an increased performance \cite{belkin2019}, both in natural language processing (e.g. \cite{kaplan2020}) and image processing tasks (e.g. \cite{zhai2022}). The second is to build stronger assumptions into the model architecture to guide its learning, and has specifically succeeded for models built on limited data (e.g. \cite{batzner2022}). The latter is the approach used for the NFST. 

These assumptions, known as inductive biases, give the model far greater generalizability for the same training data by leveraging prior knowledge or assumptions about the task and data. They limit the parameter space of the model which guides the optimization towards better solutions. They also smooth the loss landscape, promoting faster convergence and preventing the optimization from getting stuck in local minima. Inductive biases can be particularly valuable when analysing the galaxy distribution, because it has strong natural symmetries that can be leveraged: translational symmetry stemming from homogeneity and rotational symmetry stemming from isotropy. The NFST utilizes these and other inductive biases to help the optimization process converge on a more informative and robust model. 

Both the NFST and CNN models utilize convolutions as an inductive bias, stemming from the assumption of translational symmetry and invariance to small local distortions. There are differences in their implementations (Fourier space vs. configuration space) but both architectures take advantage of translational symmetry. Both models also use pooling, another inductive bias that arises from the assumption of translational symmetry. The NFST uses a strong full-field average pooling and rotational average pooling, compared to CNN models which usually utilizes more localized pooling. The NFST's stronger symmetry-based inductive biases more closely follow our understanding of the symmetries in the galaxy distribution, particularly homogeneity and isotropy. 

We note that the NFST also has an additional inductive bias in its neural field representation of the convolution kernel. This neural field imposes a functional prior, since the filter can only take forms that can be produced by the neural field. For a finitely parameterized neural field, this imposes a limit on filter complexity corresponding to the expressiveness of the neural field's network. It also fundamentally changes the loss landscape since the modification of a parameter in the NFN has an indirect effect on the frequency processing in the image, rather than a direct effect as it would have with a pixel-wise parameterized filter.

\subsection{Interpretability}
Model interpretability can allow us to distinguish between a true detection of parity-odd physical mechanisms in the data, and a false detection from any spurious parity-odd observational systematics. The NFST filters can be directly interpreted in their natural Fourier space, where a higher magnitude at a specific Fourier mode indicates that those frequencies in the data are utilized more for the detection. Alternatively, displaying the filters in configuration space shows the physical structures that influence the detection. In our tests, the NFST learns filters that directly correspond to the side lengths of the parity violating triangles. With realistic data, the structures may not be as clearly recognizable as this simplified example, but they will still provide information on which Fourier modes are most used in the detection of parity violation. 

Once an NFST model has been trained to detect parity violation, model interpretation techniques such as Shapley values \citep{lundberg2017} can also highlight the particular scattering coefficients that contribute most to the detection. Any particular NFST scattering coefficient can be directly related to at most two filters, both of which have clear associated scales and angles. This contrasts with a CNN, where every layer combines the contributions of many filters (100 filters per layer for the CNN model used in this work), for which scale and angle information is mixed in a way that is much more difficult to track. \cite{cheng2021guide} give a further discussion of the interpretability of the WST coefficients, much of which applies directly to the NFST.


\section{Conclusions}\label{sec:conclusions}
The search for parity violation in the galaxy distribution has recently garnered significant interest, due to its implications within the standard cosmological model. The typical approach to detecting parity violation involves accurately specifying the 4-point correlation function and its noise properties, which requires large suites of accurate mock simulations to estimate, presenting a significant computational and modelling challenge. We introduce a complementary method to search for parity violations, leveraging an unsupervised learning approach to provide a mock-free detection of parity violation in the galaxy distribution. The model's goal is to construct a statistic from the structures in the galaxy distribution that distinguishes between the distributions original configuration and its mirror image counterpart, thereby identifying the parity violating features. In this work, we highlight its potential in a proof-of-concept test with a simplified, 2D dataset. 

In this unsupervised model, we require a compression of the field that accesses higher-order information. In a companion paper, we utilized a convolutional neural network (CNN) model to successfully detect parity violation. However, we find the CNN struggles when the data is limited. A more robust alternative is the wavelet scattering transform (WST), a model that includes the iterative convolution properties of the CNN without any trainable parameters, with a strong set of inductive biases that are well-suited to the homogeneous and isotropic galaxy distribution. However, while the WST extracts a wealth of information, this information is not well-suited for identifying parity asymmetry. 

To address this problem, we introduce the novel Neural Field Scattering Transform (NFST), an extension to the WST that overcomes this challenge by allowing the WST filters to vary. The filters are parameterized as a neural field in Fourier space, which means the structure of the filter is embedded on the weights of a neural field network (NFN) that generates the filter as output. We demonstrate the NFST's ability to detect parity violation within the unsupervised learning framework. The main results are as follows:
\begin{enumerate}
    \item The unsupervised learning framework, when paired with an effective field compression, is a powerful approach that can be used to detect parity violation in a field. 
    \item The NFST performs best when the NFN has a width of 256 neurons, with fewer and more neurons degrading performance.
    \item Initialising the NFN to match Morlet filter structure sees an almost twofold improvement over random initialization for an NFN width of 256 neurons.
    \item Using parity asymmetric filters generally yields over $3\times$ improvement over its symmetric counterparts, indicates asymmetric filters are an important adaptation of the NFST over the standard WST. 
    \item The NFST detects parity violation with only 50 training and validation patches, requiring $32\times$ fewer training samples to detect parity violation than the WST, and $4\times$ fewer training samples than the benchmark CNN model. 
    \item The NFST can detect parity violation in the cases of 50 and 100 training and validation samples, for which the CNN and WST make no detection. The NFST's ability to find parity violation where other models cannot make it a more suitable model to use moving forward as we develop models for more realistic 3D data. 
\end{enumerate}
These results have implications for future searches for parity violation in the galaxy distribution. Unlike previous approaches, unsupervised learning can provide a detection of parity violation without requiring any simulated mock universes, making it an excellent complementary method to confirm parity violation in future surveys such as DESI and Euclid. The NFST also has valuable interpretation properties that may help disentangle systematics from signal in these future data applications. We are currently extending this method to 3D and will apply it to the search for parity violation in observational data in coming work. Overall, we demonstrate a powerful new approach that combines unsupervised learning and the novel NFST technique to detect parity violation in the galaxy distribution. 


\section{Acknowledgements}
 M.C., R.R., T.M.D., acknowledge the support of an Australian Research Council Australian Laureate Fellowship (FL180100168) funded by the Australian Government. PLT is supported in part by NASA ROSES 21-ATP21-0050. CCL is supported by the National Science Foundation under Cooperative Agreement PHY2019786 (The NSF AI Institute for Artificial Intelligence and Fundamental Interactions).This work received support from the U.S. Department of Energy under contract number DE-SC0011726. Y.S.T. acknowledges financial support from the Australian Research Council through DECRA Fellowship DE220101520. The authors thank Zachary Slepian for enlightening conversations. This research used resources of the National Energy Research Scientific Computing Center (NERSC), a U.S. Department of Energy Office of Science User Facility located at Lawrence Berkeley National Laboratory, operated under Contract No. DE-AC02-05CH11231.

\bibliographystyle{apsrev4-1.bst}
\bibliography{bibliography.bib}

\begin{thebibliography}{32}%
\makeatletter
\providecommand \@ifxundefined [1]{%
 \@ifx{#1\undefined}
}%
\providecommand \@ifnum [1]{%
 \ifnum #1\expandafter \@firstoftwo
 \else \expandafter \@secondoftwo
 \fi
}%
\providecommand \@ifx [1]{%
 \ifx #1\expandafter \@firstoftwo
 \else \expandafter \@secondoftwo
 \fi
}%
\providecommand \natexlab [1]{#1}%
\providecommand \enquote  [1]{``#1''}%
\providecommand \bibnamefont  [1]{#1}%
\providecommand \bibfnamefont [1]{#1}%
\providecommand \citenamefont [1]{#1}%
\providecommand \href@noop [0]{\@secondoftwo}%
\providecommand \href [0]{\begingroup \@sanitize@url \@href}%
\providecommand \@href[1]{\@@startlink{#1}\@@href}%
\providecommand \@@href[1]{\endgroup#1\@@endlink}%
\providecommand \@sanitize@url [0]{\catcode `\\12\catcode `\$12\catcode `\&12\catcode `\#12\catcode `\^12\catcode `\_12\catcode `\%12\relax}%
\providecommand \@@startlink[1]{}%
\providecommand \@@endlink[0]{}%
\providecommand \url  [0]{\begingroup\@sanitize@url \@url }%
\providecommand \@url [1]{\endgroup\@href {#1}{\urlprefix }}%
\providecommand \urlprefix  [0]{URL }%
\providecommand \Eprint [0]{\href }%
\providecommand \doibase [0]{http://dx.doi.org/}%
\providecommand \selectlanguage [0]{\@gobble}%
\providecommand \bibinfo  [0]{\@secondoftwo}%
\providecommand \bibfield  [0]{\@secondoftwo}%
\providecommand \translation [1]{[#1]}%
\providecommand \BibitemOpen [0]{}%
\providecommand \bibitemStop [0]{}%
\providecommand \bibitemNoStop [0]{.\EOS\space}%
\providecommand \EOS [0]{\spacefactor3000\relax}%
\providecommand \BibitemShut  [1]{\csname bibitem#1\endcsname}%
\let\auto@bib@innerbib\@empty
\bibitem [{\citenamefont {Wu}\ \emph {et~al.}(1957)\citenamefont {Wu}, \citenamefont {Ambler}, \citenamefont {Hayward}, \citenamefont {Hoppes},\ and\ \citenamefont {Hudson}}]{wu1957}%
  \BibitemOpen
  \bibfield  {author} {\bibinfo {author} {\bibfnamefont {C.~S.}\ \bibnamefont {Wu}}, \bibinfo {author} {\bibfnamefont {E.}~\bibnamefont {Ambler}}, \bibinfo {author} {\bibfnamefont {R.~W.}\ \bibnamefont {Hayward}}, \bibinfo {author} {\bibfnamefont {D.~D.}\ \bibnamefont {Hoppes}}, \ and\ \bibinfo {author} {\bibfnamefont {R.~P.}\ \bibnamefont {Hudson}},\ }\href {\doibase 10.1103/PhysRev.105.1413} {\bibfield  {journal} {\bibinfo  {journal} {Phys. Rev.}\ }\textbf {\bibinfo {volume} {105}},\ \bibinfo {pages} {1413} (\bibinfo {year} {1957})}\BibitemShut {NoStop}%
\bibitem [{\citenamefont {Schmidt}\ \emph {et~al.}(2015)\citenamefont {Schmidt}, \citenamefont {Chisari},\ and\ \citenamefont {Dvorkin}}]{schmidt2015}%
  \BibitemOpen
  \bibfield  {author} {\bibinfo {author} {\bibfnamefont {F.}~\bibnamefont {Schmidt}}, \bibinfo {author} {\bibfnamefont {N.~E.}\ \bibnamefont {Chisari}}, \ and\ \bibinfo {author} {\bibfnamefont {C.}~\bibnamefont {Dvorkin}},\ }\href {\doibase 10.1088/1475-7516/2015/10/032} {\bibfield  {journal} {\bibinfo  {journal} {Journal of Cosmology and Astroparticle Physics}\ }\textbf {\bibinfo {volume} {2015}},\ \bibinfo {pages} {032–032} (\bibinfo {year} {2015})}\BibitemShut {NoStop}%
\bibitem [{\citenamefont {Jazayeri}\ \emph {et~al.}(2023)\citenamefont {Jazayeri}, \citenamefont {Renaux-Petel}, \citenamefont {Tong}, \citenamefont {Werth},\ and\ \citenamefont {Zhu}}]{jazayeri2023}%
  \BibitemOpen
  \bibfield  {author} {\bibinfo {author} {\bibfnamefont {S.}~\bibnamefont {Jazayeri}}, \bibinfo {author} {\bibfnamefont {S.}~\bibnamefont {Renaux-Petel}}, \bibinfo {author} {\bibfnamefont {X.}~\bibnamefont {Tong}}, \bibinfo {author} {\bibfnamefont {D.}~\bibnamefont {Werth}}, \ and\ \bibinfo {author} {\bibfnamefont {Y.}~\bibnamefont {Zhu}},\ }\href@noop {} {\enquote {\bibinfo {title} {Parity violation from emergent non-locality during inflation},}\ } (\bibinfo {year} {2023}),\ \Eprint {http://arxiv.org/abs/2308.11315} {arXiv:2308.11315 [hep-th]} \BibitemShut {NoStop}%
\bibitem [{\citenamefont {Cabass}\ \emph {et~al.}(2023)\citenamefont {Cabass}, \citenamefont {Ivanov},\ and\ \citenamefont {Philcox}}]{cabass2023}%
  \BibitemOpen
  \bibfield  {author} {\bibinfo {author} {\bibfnamefont {G.}~\bibnamefont {Cabass}}, \bibinfo {author} {\bibfnamefont {M.~M.}\ \bibnamefont {Ivanov}}, \ and\ \bibinfo {author} {\bibfnamefont {O.~H.}\ \bibnamefont {Philcox}},\ }\href {\doibase 10.1103/physrevd.107.023523} {\bibfield  {journal} {\bibinfo  {journal} {Physical Review D}\ }\textbf {\bibinfo {volume} {107}},\ \bibinfo {pages} {023523} (\bibinfo {year} {2023})}\BibitemShut {NoStop}%
\bibitem [{\citenamefont {{Cahn}}\ \emph {et~al.}(2023)\citenamefont {{Cahn}}, \citenamefont {{Slepian}},\ and\ \citenamefont {{Hou}}}]{cahn2022}%
  \BibitemOpen
  \bibfield  {author} {\bibinfo {author} {\bibfnamefont {R.~N.}\ \bibnamefont {{Cahn}}}, \bibinfo {author} {\bibfnamefont {Z.}~\bibnamefont {{Slepian}}}, \ and\ \bibinfo {author} {\bibfnamefont {J.}~\bibnamefont {{Hou}}},\ }\href {\doibase 10.1103/physrevd.105.103534} {\bibfield  {journal} {\bibinfo  {journal} {\prl}\ }\textbf {\bibinfo {volume} {130}} (\bibinfo {year} {2023}),\ 10.1103/physrevd.105.103534}\BibitemShut {NoStop}%
\bibitem [{\citenamefont {{Hou}}\ \emph {et~al.}(2023)\citenamefont {{Hou}}, \citenamefont {{Slepian}},\ and\ \citenamefont {{Cahn}}}]{hou2023}%
  \BibitemOpen
  \bibfield  {author} {\bibinfo {author} {\bibfnamefont {J.}~\bibnamefont {{Hou}}}, \bibinfo {author} {\bibfnamefont {Z.}~\bibnamefont {{Slepian}}}, \ and\ \bibinfo {author} {\bibfnamefont {R.~N.}\ \bibnamefont {{Cahn}}},\ }\href {\doibase 10.1093/mnras/stad1062} {\bibfield  {journal} {\bibinfo  {journal} {Monthly Notices of the Royal Astronomical Society}\ }\textbf {\bibinfo {volume} {522}},\ \bibinfo {pages} {5701} (\bibinfo {year} {2023})},\ \Eprint {http://arxiv.org/abs/2206.03625} {arXiv:2206.03625 [astro-ph.CO]} \BibitemShut {NoStop}%
\bibitem [{\citenamefont {{Philcox}}(2022)}]{philcox2022}%
  \BibitemOpen
  \bibfield  {author} {\bibinfo {author} {\bibfnamefont {O.~H.~E.}\ \bibnamefont {{Philcox}}},\ }\href {\doibase 10.1103/PhysRevD.106.063501} {\bibfield  {journal} {\bibinfo  {journal} {\prd}\ }\textbf {\bibinfo {volume} {106}},\ \bibinfo {eid} {063501} (\bibinfo {year} {2022})},\ \Eprint {http://arxiv.org/abs/2206.04227} {arXiv:2206.04227 [astro-ph.CO]} \BibitemShut {NoStop}%
\bibitem [{\citenamefont {Philcox}\ and\ \citenamefont {Ereza}(2024)}]{philcox2024}%
  \BibitemOpen
  \bibfield  {author} {\bibinfo {author} {\bibfnamefont {O.~H.~E.}\ \bibnamefont {Philcox}}\ and\ \bibinfo {author} {\bibfnamefont {J.}~\bibnamefont {Ereza}},\ }\href@noop {} {\enquote {\bibinfo {title} {Could sample variance be responsible for the parity-violating signal seen in the boss galaxy survey?}}\ } (\bibinfo {year} {2024}),\ \Eprint {http://arxiv.org/abs/2401.09523} {arXiv:2401.09523 [astro-ph.CO]} \BibitemShut {NoStop}%
\bibitem [{\citenamefont {Ereza}\ \emph {et~al.}(2023)\citenamefont {Ereza}, \citenamefont {Prada}, \citenamefont {Klypin}, \citenamefont {Ishiyama}, \citenamefont {Smith}, \citenamefont {Baugh}, \citenamefont {Li}, \citenamefont {Hernández-Aguayo},\ and\ \citenamefont {Ruedas}}]{ereza2023}%
  \BibitemOpen
  \bibfield  {author} {\bibinfo {author} {\bibfnamefont {J.}~\bibnamefont {Ereza}}, \bibinfo {author} {\bibfnamefont {F.}~\bibnamefont {Prada}}, \bibinfo {author} {\bibfnamefont {A.}~\bibnamefont {Klypin}}, \bibinfo {author} {\bibfnamefont {T.}~\bibnamefont {Ishiyama}}, \bibinfo {author} {\bibfnamefont {A.}~\bibnamefont {Smith}}, \bibinfo {author} {\bibfnamefont {C.~M.}\ \bibnamefont {Baugh}}, \bibinfo {author} {\bibfnamefont {B.}~\bibnamefont {Li}}, \bibinfo {author} {\bibfnamefont {C.}~\bibnamefont {Hernández-Aguayo}}, \ and\ \bibinfo {author} {\bibfnamefont {J.}~\bibnamefont {Ruedas}},\ }\href@noop {} {\enquote {\bibinfo {title} {The uchuu-glam boss and eboss lrg lightcones: Exploring clustering and covariance errors},}\ } (\bibinfo {year} {2023}),\ \Eprint {http://arxiv.org/abs/2311.14456} {arXiv:2311.14456 [astro-ph.CO]} \BibitemShut {NoStop}%
\bibitem [{\citenamefont {Rodríguez-Torres}\ \emph {et~al.}(2016)\citenamefont {Rodríguez-Torres}, \citenamefont {Chuang}, \citenamefont {Prada}, \citenamefont {Guo}, \citenamefont {Klypin}, \citenamefont {Behroozi}, \citenamefont {Hahn}, \citenamefont {Comparat}, \citenamefont {Yepes}, \citenamefont {Montero-Dorta}, \citenamefont {Brownstein}, \citenamefont {Maraston}, \citenamefont {McBride}, \citenamefont {Tinker}, \citenamefont {Gottlöber}, \citenamefont {Favole}, \citenamefont {Shu}, \citenamefont {Kitaura}, \citenamefont {Bolton}, \citenamefont {Scoccimarro}, \citenamefont {Samushia}, \citenamefont {Schlegel}, \citenamefont {Schneider},\ and\ \citenamefont {Thomas}}]{rodrigueztorres2016}%
  \BibitemOpen
  \bibfield  {author} {\bibinfo {author} {\bibfnamefont {S.~A.}\ \bibnamefont {Rodríguez-Torres}}, \bibinfo {author} {\bibfnamefont {C.-H.}\ \bibnamefont {Chuang}}, \bibinfo {author} {\bibfnamefont {F.}~\bibnamefont {Prada}}, \bibinfo {author} {\bibfnamefont {H.}~\bibnamefont {Guo}}, \bibinfo {author} {\bibfnamefont {A.}~\bibnamefont {Klypin}}, \bibinfo {author} {\bibfnamefont {P.}~\bibnamefont {Behroozi}}, \bibinfo {author} {\bibfnamefont {C.~H.}\ \bibnamefont {Hahn}}, \bibinfo {author} {\bibfnamefont {J.}~\bibnamefont {Comparat}}, \bibinfo {author} {\bibfnamefont {G.}~\bibnamefont {Yepes}}, \bibinfo {author} {\bibfnamefont {A.~D.}\ \bibnamefont {Montero-Dorta}}, \bibinfo {author} {\bibfnamefont {J.~R.}\ \bibnamefont {Brownstein}}, \bibinfo {author} {\bibfnamefont {C.}~\bibnamefont {Maraston}}, \bibinfo {author} {\bibfnamefont {C.~K.}\ \bibnamefont {McBride}}, \bibinfo {author} {\bibfnamefont {J.}~\bibnamefont {Tinker}}, \bibinfo {author} {\bibfnamefont {S.}~\bibnamefont {Gottlöber}}, \bibinfo {author}
  {\bibfnamefont {G.}~\bibnamefont {Favole}}, \bibinfo {author} {\bibfnamefont {Y.}~\bibnamefont {Shu}}, \bibinfo {author} {\bibfnamefont {F.-S.}\ \bibnamefont {Kitaura}}, \bibinfo {author} {\bibfnamefont {A.}~\bibnamefont {Bolton}}, \bibinfo {author} {\bibfnamefont {R.}~\bibnamefont {Scoccimarro}}, \bibinfo {author} {\bibfnamefont {L.}~\bibnamefont {Samushia}}, \bibinfo {author} {\bibfnamefont {D.}~\bibnamefont {Schlegel}}, \bibinfo {author} {\bibfnamefont {D.~P.}\ \bibnamefont {Schneider}}, \ and\ \bibinfo {author} {\bibfnamefont {D.}~\bibnamefont {Thomas}},\ }\href {\doibase 10.1093/mnras/stw1014} {\bibfield  {journal} {\bibinfo  {journal} {Monthly Notices of the Royal Astronomical Society}\ }\textbf {\bibinfo {volume} {460}},\ \bibinfo {pages} {1173–1187} (\bibinfo {year} {2016})}\BibitemShut {NoStop}%
\bibitem [{\citenamefont {Kitaura}\ \emph {et~al.}(2016)\citenamefont {Kitaura}, \citenamefont {Rodríguez-Torres}, \citenamefont {Chuang}, \citenamefont {Zhao}, \citenamefont {Prada}, \citenamefont {Gil-Marín}, \citenamefont {Guo}, \citenamefont {Yepes}, \citenamefont {Klypin}, \citenamefont {Scóccola}, \citenamefont {Tinker}, \citenamefont {McBride}, \citenamefont {Reid}, \citenamefont {Sánchez}, \citenamefont {Salazar-Albornoz}, \citenamefont {Grieb}, \citenamefont {Vargas-Magana}, \citenamefont {Cuesta}, \citenamefont {Neyrinck}, \citenamefont {Beutler}, \citenamefont {Comparat}, \citenamefont {Percival},\ and\ \citenamefont {Ross}}]{kitaura2016}%
  \BibitemOpen
  \bibfield  {author} {\bibinfo {author} {\bibfnamefont {F.-S.}\ \bibnamefont {Kitaura}}, \bibinfo {author} {\bibfnamefont {S.}~\bibnamefont {Rodríguez-Torres}}, \bibinfo {author} {\bibfnamefont {C.-H.}\ \bibnamefont {Chuang}}, \bibinfo {author} {\bibfnamefont {C.}~\bibnamefont {Zhao}}, \bibinfo {author} {\bibfnamefont {F.}~\bibnamefont {Prada}}, \bibinfo {author} {\bibfnamefont {H.}~\bibnamefont {Gil-Marín}}, \bibinfo {author} {\bibfnamefont {H.}~\bibnamefont {Guo}}, \bibinfo {author} {\bibfnamefont {G.}~\bibnamefont {Yepes}}, \bibinfo {author} {\bibfnamefont {A.}~\bibnamefont {Klypin}}, \bibinfo {author} {\bibfnamefont {C.~G.}\ \bibnamefont {Scóccola}}, \bibinfo {author} {\bibfnamefont {J.}~\bibnamefont {Tinker}}, \bibinfo {author} {\bibfnamefont {C.}~\bibnamefont {McBride}}, \bibinfo {author} {\bibfnamefont {B.}~\bibnamefont {Reid}}, \bibinfo {author} {\bibfnamefont {A.~G.}\ \bibnamefont {Sánchez}}, \bibinfo {author} {\bibfnamefont {S.}~\bibnamefont {Salazar-Albornoz}}, \bibinfo {author} {\bibfnamefont
  {J.~N.}\ \bibnamefont {Grieb}}, \bibinfo {author} {\bibfnamefont {M.}~\bibnamefont {Vargas-Magana}}, \bibinfo {author} {\bibfnamefont {A.~J.}\ \bibnamefont {Cuesta}}, \bibinfo {author} {\bibfnamefont {M.}~\bibnamefont {Neyrinck}}, \bibinfo {author} {\bibfnamefont {F.}~\bibnamefont {Beutler}}, \bibinfo {author} {\bibfnamefont {J.}~\bibnamefont {Comparat}}, \bibinfo {author} {\bibfnamefont {W.~J.}\ \bibnamefont {Percival}}, \ and\ \bibinfo {author} {\bibfnamefont {A.}~\bibnamefont {Ross}},\ }\href {\doibase 10.1093/mnras/stv2826} {\bibfield  {journal} {\bibinfo  {journal} {Monthly Notices of the Royal Astronomical Society}\ }\textbf {\bibinfo {volume} {456}},\ \bibinfo {pages} {4156–4173} (\bibinfo {year} {2016})}\BibitemShut {NoStop}%
\bibitem [{\citenamefont {Lester}\ and\ \citenamefont {Tombs}(2022)}]{lester2022a}%
  \BibitemOpen
  \bibfield  {author} {\bibinfo {author} {\bibfnamefont {C.~G.}\ \bibnamefont {Lester}}\ and\ \bibinfo {author} {\bibfnamefont {R.}~\bibnamefont {Tombs}},\ }\href@noop {} {\enquote {\bibinfo {title} {Using unsupervised learning to detect broken symmetries, with relevance to searches for parity violation in nature. (previously: "stressed gans snag desserts")},}\ } (\bibinfo {year} {2022}),\ \Eprint {http://arxiv.org/abs/2111.00616} {arXiv:2111.00616 [hep-ph]} \BibitemShut {NoStop}%
\bibitem [{\citenamefont {Lester}\ \emph {et~al.}(2022)\citenamefont {Lester}, \citenamefont {Mastandrea}, \citenamefont {Noel},\ and\ \citenamefont {Tombs}}]{lester2022b}%
  \BibitemOpen
  \bibfield  {author} {\bibinfo {author} {\bibfnamefont {C.~G.}\ \bibnamefont {Lester}}, \bibinfo {author} {\bibfnamefont {R.}~\bibnamefont {Mastandrea}}, \bibinfo {author} {\bibfnamefont {D.}~\bibnamefont {Noel}}, \ and\ \bibinfo {author} {\bibfnamefont {R.}~\bibnamefont {Tombs}},\ }\href {\doibase 10.1007/jhep08(2022)231} {\bibfield  {journal} {\bibinfo  {journal} {Journal of High Energy Physics}\ }\textbf {\bibinfo {volume} {2022}} (\bibinfo {year} {2022}),\ 10.1007/jhep08(2022)231}\BibitemShut {NoStop}%
\bibitem [{\citenamefont {Tombs}\ and\ \citenamefont {Lester}(2022)}]{tombs2022}%
  \BibitemOpen
  \bibfield  {author} {\bibinfo {author} {\bibfnamefont {R.}~\bibnamefont {Tombs}}\ and\ \bibinfo {author} {\bibfnamefont {C.~G.}\ \bibnamefont {Lester}},\ }\href {\doibase 10.1088/1748-0221/17/08/p08024} {\bibfield  {journal} {\bibinfo  {journal} {Journal of Instrumentation}\ }\textbf {\bibinfo {volume} {17}},\ \bibinfo {pages} {P08024} (\bibinfo {year} {2022})}\BibitemShut {NoStop}%
\bibitem [{\citenamefont {Taylor}\ \emph {et~al.}(2024)\citenamefont {Taylor}, \citenamefont {Craigie},\ and\ \citenamefont {Ting}}]{taylor2023}%
  \BibitemOpen
  \bibfield  {author} {\bibinfo {author} {\bibfnamefont {P.~L.}\ \bibnamefont {Taylor}}, \bibinfo {author} {\bibfnamefont {M.}~\bibnamefont {Craigie}}, \ and\ \bibinfo {author} {\bibfnamefont {Y.-S.}\ \bibnamefont {Ting}},\ }\href {\doibase 10.1103/PhysRevD.109.083518} {\bibfield  {journal} {\bibinfo  {journal} {Phys. Rev. D}\ }\textbf {\bibinfo {volume} {109}},\ \bibinfo {pages} {083518} (\bibinfo {year} {2024})}\BibitemShut {NoStop}%
\bibitem [{\citenamefont {Mallat}(2012)}]{mallat2012}%
  \BibitemOpen
  \bibfield  {author} {\bibinfo {author} {\bibfnamefont {S.}~\bibnamefont {Mallat}},\ }\href {\doibase https://doi.org/10.1002/cpa.21413} {\bibfield  {journal} {\bibinfo  {journal} {Communications on Pure and Applied Mathematics}\ }\textbf {\bibinfo {volume} {65}},\ \bibinfo {pages} {1331} (\bibinfo {year} {2012})}\BibitemShut {NoStop}%
\bibitem [{\citenamefont {Valogiannis}\ and\ \citenamefont {Dvorkin}(2022)}]{valogiannis2022a}%
  \BibitemOpen
  \bibfield  {author} {\bibinfo {author} {\bibfnamefont {G.}~\bibnamefont {Valogiannis}}\ and\ \bibinfo {author} {\bibfnamefont {C.}~\bibnamefont {Dvorkin}},\ }\href {\doibase 10.1103/physrevd.105.103534} {\bibfield  {journal} {\bibinfo  {journal} {Physical Review D}\ }\textbf {\bibinfo {volume} {105}},\ \bibinfo {pages} {103534} (\bibinfo {year} {2022})}\BibitemShut {NoStop}%
\bibitem [{\citenamefont {{Valogiannis}}\ \emph {et~al.}(2023)\citenamefont {{Valogiannis}}, \citenamefont {{Yuan}},\ and\ \citenamefont {{Dvorkin}}}]{valogiannis2023}%
  \BibitemOpen
  \bibfield  {author} {\bibinfo {author} {\bibfnamefont {G.}~\bibnamefont {{Valogiannis}}}, \bibinfo {author} {\bibfnamefont {S.}~\bibnamefont {{Yuan}}}, \ and\ \bibinfo {author} {\bibfnamefont {C.}~\bibnamefont {{Dvorkin}}},\ }\href {\doibase 10.48550/arXiv.2310.16116} {\bibfield  {journal} {\bibinfo  {journal} {arXiv e-prints}\ ,\ \bibinfo {eid} {arXiv:2310.16116}} (\bibinfo {year} {2023})},\ \Eprint {http://arxiv.org/abs/2310.16116} {arXiv:2310.16116 [astro-ph.CO]} \BibitemShut {NoStop}%
\bibitem [{\citenamefont {{R{\'e}galdo-Saint Blancard}}\ \emph {et~al.}(2023)\citenamefont {{R{\'e}galdo-Saint Blancard}}, \citenamefont {{Hahn}}, \citenamefont {{Ho}}, \citenamefont {{Hou}}, \citenamefont {{Lemos}}, \citenamefont {{Massara}}, \citenamefont {{Modi}}, \citenamefont {{Moradinezhad Dizgah}}, \citenamefont {{Parker}}, \citenamefont {{Yao}},\ and\ \citenamefont {{Eickenberg}}}]{regaldosaintblancard2023}%
  \BibitemOpen
  \bibfield  {author} {\bibinfo {author} {\bibfnamefont {B.}~\bibnamefont {{R{\'e}galdo-Saint Blancard}}}, \bibinfo {author} {\bibfnamefont {C.}~\bibnamefont {{Hahn}}}, \bibinfo {author} {\bibfnamefont {S.}~\bibnamefont {{Ho}}}, \bibinfo {author} {\bibfnamefont {J.}~\bibnamefont {{Hou}}}, \bibinfo {author} {\bibfnamefont {P.}~\bibnamefont {{Lemos}}}, \bibinfo {author} {\bibfnamefont {E.}~\bibnamefont {{Massara}}}, \bibinfo {author} {\bibfnamefont {C.}~\bibnamefont {{Modi}}}, \bibinfo {author} {\bibfnamefont {A.}~\bibnamefont {{Moradinezhad Dizgah}}}, \bibinfo {author} {\bibfnamefont {L.}~\bibnamefont {{Parker}}}, \bibinfo {author} {\bibfnamefont {Y.}~\bibnamefont {{Yao}}}, \ and\ \bibinfo {author} {\bibfnamefont {M.}~\bibnamefont {{Eickenberg}}},\ }\href {\doibase 10.48550/arXiv.2310.15250} {\bibfield  {journal} {\bibinfo  {journal} {arXiv e-prints}\ ,\ \bibinfo {eid} {arXiv:2310.15250}} (\bibinfo {year} {2023})},\ \Eprint {http://arxiv.org/abs/2310.15250} {arXiv:2310.15250 [astro-ph.CO]} \BibitemShut
  {NoStop}%
\bibitem [{\citenamefont {Valogiannis}\ \emph {et~al.}(2024)\citenamefont {Valogiannis}, \citenamefont {Yuan},\ and\ \citenamefont {Dvorkin}}]{valogiannis2024}%
  \BibitemOpen
  \bibfield  {author} {\bibinfo {author} {\bibfnamefont {G.}~\bibnamefont {Valogiannis}}, \bibinfo {author} {\bibfnamefont {S.}~\bibnamefont {Yuan}}, \ and\ \bibinfo {author} {\bibfnamefont {C.}~\bibnamefont {Dvorkin}},\ }\href@noop {} {\enquote {\bibinfo {title} {Precise cosmological constraints from boss galaxy clustering with a simulation-based emulator of the wavelet scattering transform},}\ } (\bibinfo {year} {2024}),\ \Eprint {http://arxiv.org/abs/2310.16116} {arXiv:2310.16116 [astro-ph.CO]} \BibitemShut {NoStop}%
\bibitem [{\citenamefont {Cheng}\ and\ \citenamefont {Ménard}(2021)}]{cheng2021guide}%
  \BibitemOpen
  \bibfield  {author} {\bibinfo {author} {\bibfnamefont {S.}~\bibnamefont {Cheng}}\ and\ \bibinfo {author} {\bibfnamefont {B.}~\bibnamefont {Ménard}},\ }\href {\doibase 10.48550/ARXIV.2112.01288} {\bibfield  {journal} {\bibinfo  {journal} {arXiv e-prints}\ } (\bibinfo {year} {2021}),\ 10.48550/ARXIV.2112.01288}\BibitemShut {NoStop}%
\bibitem [{\citenamefont {{Morlet}}\ \emph {et~al.}(1982)\citenamefont {{Morlet}}, \citenamefont {{Arens}}, \citenamefont {{Forgeau}},\ and\ \citenamefont {{Giard}}}]{morlet1982}%
  \BibitemOpen
  \bibfield  {author} {\bibinfo {author} {\bibfnamefont {J.}~\bibnamefont {{Morlet}}}, \bibinfo {author} {\bibfnamefont {G.}~\bibnamefont {{Arens}}}, \bibinfo {author} {\bibfnamefont {I.}~\bibnamefont {{Forgeau}}}, \ and\ \bibinfo {author} {\bibfnamefont {D.}~\bibnamefont {{Giard}}},\ }\href {\doibase 10.1190/1.1441328} {\bibfield  {journal} {\bibinfo  {journal} {Geophysics}\ }\textbf {\bibinfo {volume} {47}},\ \bibinfo {pages} {203} (\bibinfo {year} {1982})}\BibitemShut {NoStop}%
\bibitem [{\citenamefont {{Xie}}\ \emph {et~al.}(2022)\citenamefont {{Xie}}, \citenamefont {{Takikawa}}, \citenamefont {{Saito}}, \citenamefont {{Litany}}, \citenamefont {{Yan}}, \citenamefont {{Khan}}, \citenamefont {{Tombari}}, \citenamefont {{Tompkin}}, \citenamefont {{Sitzmann}},\ and\ \citenamefont {{Sridhar}}}]{xie2021}%
  \BibitemOpen
  \bibfield  {author} {\bibinfo {author} {\bibfnamefont {Y.}~\bibnamefont {{Xie}}}, \bibinfo {author} {\bibfnamefont {T.}~\bibnamefont {{Takikawa}}}, \bibinfo {author} {\bibfnamefont {S.}~\bibnamefont {{Saito}}}, \bibinfo {author} {\bibfnamefont {O.}~\bibnamefont {{Litany}}}, \bibinfo {author} {\bibfnamefont {S.}~\bibnamefont {{Yan}}}, \bibinfo {author} {\bibfnamefont {N.}~\bibnamefont {{Khan}}}, \bibinfo {author} {\bibfnamefont {F.}~\bibnamefont {{Tombari}}}, \bibinfo {author} {\bibfnamefont {J.}~\bibnamefont {{Tompkin}}}, \bibinfo {author} {\bibfnamefont {V.}~\bibnamefont {{Sitzmann}}}, \ and\ \bibinfo {author} {\bibfnamefont {S.}~\bibnamefont {{Sridhar}}},\ }\href {\doibase https://doi.org/10.1111/cgf.14505} {\bibfield  {journal} {\bibinfo  {journal} {Computer Graphics Forum}\ ,\ \bibinfo {pages} {641}} (\bibinfo {year} {2022})}\BibitemShut {NoStop}%
\bibitem [{\citenamefont {Mildenhall}\ \emph {et~al.}(2021)\citenamefont {Mildenhall}, \citenamefont {Srinivasan}, \citenamefont {Tancik}, \citenamefont {Barron}, \citenamefont {Ramamoorthi},\ and\ \citenamefont {Ng}}]{mildenhall2020}%
  \BibitemOpen
  \bibfield  {author} {\bibinfo {author} {\bibfnamefont {B.}~\bibnamefont {Mildenhall}}, \bibinfo {author} {\bibfnamefont {P.~P.}\ \bibnamefont {Srinivasan}}, \bibinfo {author} {\bibfnamefont {M.}~\bibnamefont {Tancik}}, \bibinfo {author} {\bibfnamefont {J.~T.}\ \bibnamefont {Barron}}, \bibinfo {author} {\bibfnamefont {R.}~\bibnamefont {Ramamoorthi}}, \ and\ \bibinfo {author} {\bibfnamefont {R.}~\bibnamefont {Ng}},\ }\href {\doibase 10.1145/3503250} {\bibfield  {journal} {\bibinfo  {journal} {Commun. ACM}\ }\textbf {\bibinfo {volume} {65}},\ \bibinfo {pages} {99–106} (\bibinfo {year} {2021})}\BibitemShut {NoStop}%
\bibitem [{\citenamefont {{Allys}}\ \emph {et~al.}(2019)\citenamefont {{Allys}}, \citenamefont {{Levrier}}, \citenamefont {{Zhang}}, \citenamefont {{Colling}}, \citenamefont {{Regaldo-Saint Blancard}}, \citenamefont {{Boulanger}}, \citenamefont {{Hennebelle}},\ and\ \citenamefont {{Mallat}}}]{allys2019}%
  \BibitemOpen
  \bibfield  {author} {\bibinfo {author} {\bibfnamefont {E.}~\bibnamefont {{Allys}}}, \bibinfo {author} {\bibfnamefont {F.}~\bibnamefont {{Levrier}}}, \bibinfo {author} {\bibfnamefont {S.}~\bibnamefont {{Zhang}}}, \bibinfo {author} {\bibfnamefont {C.}~\bibnamefont {{Colling}}}, \bibinfo {author} {\bibfnamefont {B.}~\bibnamefont {{Regaldo-Saint Blancard}}}, \bibinfo {author} {\bibfnamefont {F.}~\bibnamefont {{Boulanger}}}, \bibinfo {author} {\bibfnamefont {P.}~\bibnamefont {{Hennebelle}}}, \ and\ \bibinfo {author} {\bibfnamefont {S.}~\bibnamefont {{Mallat}}},\ }\href {\doibase 10.1051/0004-6361/201834975} {\bibfield  {journal} {\bibinfo  {journal} {Astronomy and Astrophysics}\ }\textbf {\bibinfo {volume} {629}},\ \bibinfo {eid} {A115} (\bibinfo {year} {2019})},\ \Eprint {http://arxiv.org/abs/1905.01372} {arXiv:1905.01372 [astro-ph.CO]} \BibitemShut {NoStop}%
\bibitem [{\citenamefont {Krizhevsky}\ \emph {et~al.}(2012)\citenamefont {Krizhevsky}, \citenamefont {Sutskever},\ and\ \citenamefont {Hinton}}]{krizhevsky2012}%
  \BibitemOpen
  \bibfield  {author} {\bibinfo {author} {\bibfnamefont {A.}~\bibnamefont {Krizhevsky}}, \bibinfo {author} {\bibfnamefont {I.}~\bibnamefont {Sutskever}}, \ and\ \bibinfo {author} {\bibfnamefont {G.~E.}\ \bibnamefont {Hinton}},\ }in\ \href {https://proceedings.neurips.cc/paper_files/paper/2012/file/c399862d3b9d6b76c8436e924a68c45b-Paper.pdf} {\emph {\bibinfo {booktitle} {Advances in Neural Information Processing Systems}}},\ Vol.~\bibinfo {volume} {25},\ \bibinfo {editor} {edited by\ \bibinfo {editor} {\bibfnamefont {F.}~\bibnamefont {Pereira}}, \bibinfo {editor} {\bibfnamefont {C.}~\bibnamefont {Burges}}, \bibinfo {editor} {\bibfnamefont {L.}~\bibnamefont {Bottou}}, \ and\ \bibinfo {editor} {\bibfnamefont {K.}~\bibnamefont {Weinberger}}}\ (\bibinfo  {publisher} {Curran Associates, Inc.},\ \bibinfo {year} {2012})\BibitemShut {NoStop}%
\bibitem [{\citenamefont {Andreux}\ \emph {et~al.}(2020)\citenamefont {Andreux}, \citenamefont {Angles}, \citenamefont {Exarchakis}, \citenamefont {Leonarduzzi}, \citenamefont {Rochette}, \citenamefont {Thiry}, \citenamefont {Zarka}, \citenamefont {Mallat}, \citenamefont {AndÃ©n}, \citenamefont {Belilovsky}, \citenamefont {Bruna}, \citenamefont {Lostanlen}, \citenamefont {Chaudhary}, \citenamefont {Hirn}, \citenamefont {Oyallon}, \citenamefont {Zhang}, \citenamefont {Cella},\ and\ \citenamefont {Eickenberg}}]{andreux2020kymatio}%
  \BibitemOpen
  \bibfield  {author} {\bibinfo {author} {\bibfnamefont {M.}~\bibnamefont {Andreux}}, \bibinfo {author} {\bibfnamefont {T.}~\bibnamefont {Angles}}, \bibinfo {author} {\bibfnamefont {G.}~\bibnamefont {Exarchakis}}, \bibinfo {author} {\bibfnamefont {R.}~\bibnamefont {Leonarduzzi}}, \bibinfo {author} {\bibfnamefont {G.}~\bibnamefont {Rochette}}, \bibinfo {author} {\bibfnamefont {L.}~\bibnamefont {Thiry}}, \bibinfo {author} {\bibfnamefont {J.}~\bibnamefont {Zarka}}, \bibinfo {author} {\bibfnamefont {S.}~\bibnamefont {Mallat}}, \bibinfo {author} {\bibfnamefont {J.}~\bibnamefont {AndÃ©n}}, \bibinfo {author} {\bibfnamefont {E.}~\bibnamefont {Belilovsky}}, \bibinfo {author} {\bibfnamefont {J.}~\bibnamefont {Bruna}}, \bibinfo {author} {\bibfnamefont {V.}~\bibnamefont {Lostanlen}}, \bibinfo {author} {\bibfnamefont {M.}~\bibnamefont {Chaudhary}}, \bibinfo {author} {\bibfnamefont {M.~J.}\ \bibnamefont {Hirn}}, \bibinfo {author} {\bibfnamefont {E.}~\bibnamefont {Oyallon}}, \bibinfo {author} {\bibfnamefont
  {S.}~\bibnamefont {Zhang}}, \bibinfo {author} {\bibfnamefont {C.}~\bibnamefont {Cella}}, \ and\ \bibinfo {author} {\bibfnamefont {M.}~\bibnamefont {Eickenberg}},\ }\href {http://jmlr.org/papers/v21/19-047.html} {\bibfield  {journal} {\bibinfo  {journal} {Journal of Machine Learning Research}\ }\textbf {\bibinfo {volume} {21}},\ \bibinfo {pages} {1} (\bibinfo {year} {2020})}\BibitemShut {NoStop}%
\bibitem [{\citenamefont {Belkin}\ \emph {et~al.}(2019)\citenamefont {Belkin}, \citenamefont {Hsu}, \citenamefont {Ma},\ and\ \citenamefont {Mandal}}]{belkin2019}%
  \BibitemOpen
  \bibfield  {author} {\bibinfo {author} {\bibfnamefont {M.}~\bibnamefont {Belkin}}, \bibinfo {author} {\bibfnamefont {D.}~\bibnamefont {Hsu}}, \bibinfo {author} {\bibfnamefont {S.}~\bibnamefont {Ma}}, \ and\ \bibinfo {author} {\bibfnamefont {S.}~\bibnamefont {Mandal}},\ }\href {\doibase 10.1073/pnas.1903070116} {\bibfield  {journal} {\bibinfo  {journal} {Proceedings of the National Academy of Sciences}\ }\textbf {\bibinfo {volume} {116}},\ \bibinfo {pages} {15849–15854} (\bibinfo {year} {2019})}\BibitemShut {NoStop}%
\bibitem [{\citenamefont {Kaplan}\ \emph {et~al.}(2020)\citenamefont {Kaplan}, \citenamefont {McCandlish}, \citenamefont {Henighan}, \citenamefont {Brown}, \citenamefont {Chess}, \citenamefont {Child}, \citenamefont {Gray}, \citenamefont {Radford}, \citenamefont {Wu},\ and\ \citenamefont {Amodei}}]{kaplan2020}%
  \BibitemOpen
  \bibfield  {author} {\bibinfo {author} {\bibfnamefont {J.}~\bibnamefont {Kaplan}}, \bibinfo {author} {\bibfnamefont {S.}~\bibnamefont {McCandlish}}, \bibinfo {author} {\bibfnamefont {T.}~\bibnamefont {Henighan}}, \bibinfo {author} {\bibfnamefont {T.~B.}\ \bibnamefont {Brown}}, \bibinfo {author} {\bibfnamefont {B.}~\bibnamefont {Chess}}, \bibinfo {author} {\bibfnamefont {R.}~\bibnamefont {Child}}, \bibinfo {author} {\bibfnamefont {S.}~\bibnamefont {Gray}}, \bibinfo {author} {\bibfnamefont {A.}~\bibnamefont {Radford}}, \bibinfo {author} {\bibfnamefont {J.}~\bibnamefont {Wu}}, \ and\ \bibinfo {author} {\bibfnamefont {D.}~\bibnamefont {Amodei}},\ }\href@noop {} {\enquote {\bibinfo {title} {Scaling laws for neural language models},}\ } (\bibinfo {year} {2020}),\ \Eprint {http://arxiv.org/abs/2001.08361} {arXiv:2001.08361 [cs.LG]} \BibitemShut {NoStop}%
\bibitem [{\citenamefont {Zhai}\ \emph {et~al.}(2022)\citenamefont {Zhai}, \citenamefont {Kolesnikov}, \citenamefont {Houlsby},\ and\ \citenamefont {Beyer}}]{zhai2022}%
  \BibitemOpen
  \bibfield  {author} {\bibinfo {author} {\bibfnamefont {X.}~\bibnamefont {Zhai}}, \bibinfo {author} {\bibfnamefont {A.}~\bibnamefont {Kolesnikov}}, \bibinfo {author} {\bibfnamefont {N.}~\bibnamefont {Houlsby}}, \ and\ \bibinfo {author} {\bibfnamefont {L.}~\bibnamefont {Beyer}},\ }\href@noop {} {\enquote {\bibinfo {title} {Scaling vision transformers},}\ } (\bibinfo {year} {2022}),\ \Eprint {http://arxiv.org/abs/2106.04560} {arXiv:2106.04560 [cs.CV]} \BibitemShut {NoStop}%
\bibitem [{\citenamefont {Batzner}\ \emph {et~al.}(2022)\citenamefont {Batzner}, \citenamefont {Musaelian}, \citenamefont {Sun}, \citenamefont {Geiger}, \citenamefont {Mailoa}, \citenamefont {Kornbluth}, \citenamefont {Molinari}, \citenamefont {Smidt},\ and\ \citenamefont {Kozinsky}}]{batzner2022}%
  \BibitemOpen
  \bibfield  {author} {\bibinfo {author} {\bibfnamefont {S.}~\bibnamefont {Batzner}}, \bibinfo {author} {\bibfnamefont {A.}~\bibnamefont {Musaelian}}, \bibinfo {author} {\bibfnamefont {L.}~\bibnamefont {Sun}}, \bibinfo {author} {\bibfnamefont {M.}~\bibnamefont {Geiger}}, \bibinfo {author} {\bibfnamefont {J.~P.}\ \bibnamefont {Mailoa}}, \bibinfo {author} {\bibfnamefont {M.}~\bibnamefont {Kornbluth}}, \bibinfo {author} {\bibfnamefont {N.}~\bibnamefont {Molinari}}, \bibinfo {author} {\bibfnamefont {T.~E.}\ \bibnamefont {Smidt}}, \ and\ \bibinfo {author} {\bibfnamefont {B.}~\bibnamefont {Kozinsky}},\ }\href {\doibase 10.1038/s41467-022-29939-5} {\bibfield  {journal} {\bibinfo  {journal} {Nature Communications}\ }\textbf {\bibinfo {volume} {13}} (\bibinfo {year} {2022}),\ 10.1038/s41467-022-29939-5}\BibitemShut {NoStop}%
\bibitem [{\citenamefont {Lundberg}\ and\ \citenamefont {Lee}(2017)}]{lundberg2017}%
  \BibitemOpen
  \bibfield  {author} {\bibinfo {author} {\bibfnamefont {S.~M.}\ \bibnamefont {Lundberg}}\ and\ \bibinfo {author} {\bibfnamefont {S.-I.}\ \bibnamefont {Lee}},\ }in\ \href@noop {} {\emph {\bibinfo {booktitle} {Proceedings of the 31st International Conference on Neural Information Processing Systems}}},\ \bibinfo {series and number} {NIPS'17}\ (\bibinfo  {publisher} {Curran Associates Inc.},\ \bibinfo {address} {Red Hook, NY, USA},\ \bibinfo {year} {2017})\ p.\ \bibinfo {pages} {4768–4777}\BibitemShut {NoStop}%
\end{thebibliography}%

\section{Appendix}\label{sec:appendix}
\subsection{Morlet Wavelets}\label{sec:appendix_morlet}
We outline the specific details of Morlet wavelets, which are well-suited for use as the filter in the WST. In configuration space, the Morlet wavelets are complex plane waves modulated by a Gaussian kernel. The filter's direction and scale are governed by the plane wave direction and wavelength, respectively. The Gaussian modulation ensures that the convolution maintains spatial localization, which is crucial to correlate scales across repeated convolutions. This form of filters provides an optimal balance between localization in frequency and configuration spaces. The base filter in configuration space is
\begin{equation}
    \Psi(\vec{r}) = \frac{1}{\sqrt{\det\Sigma}} 
    \exp\left(-\frac{1}{2}\vec{r}^\textrm{T} \Sigma \vec{r}\right)
    \left[\exp\left(i \vec{k_0}\cdot\vec{r}\right) - \beta \right]
\end{equation}
where $\Sigma$ is the covariance matrix specifying the shape of the Gaussian kernel and $k_0$ is a vector specifying the spatial frequency and direction of the plane wave. The additional term $\beta$ is a scalar used to ensure the wavelet satisfies the admissibility criterion that it has a zero mean, and is given by
\begin{equation}
    \beta = \exp\left(-\frac{1}{2}\vec{k}_0^\textrm{T} \Sigma \vec{k}_0\right).
\end{equation}
The admissibility criterion ensures that the wavelet does not contribute to the values of the scattering coefficients when convolved. 

Often, it is easier to work with the Morlet filters in Fourier space where they are simply a real-only Gaussian window centred around a nonzero frequency. Since convolutions in configuration space are equivalent to pointwise products in Fourier space, the Morlet filters act as a directional band-pass filter. The admissibility criterion enters as a subtracted Gaussian centered at zero frequency. The base filter in Fourier space is
\begin{equation}
\begin{split}
    \hat{\Psi}(\vec{k}) = & 
    \exp\left(-\frac{1}{2}(\vec{k} - \vec{k}_0)^\textrm{T} \Sigma (\vec{k} - \vec{k}_0) \right) \\
    & - \beta \exp\left(-\frac{1}{2}\vec{k}^\textrm{T} \Sigma \vec{k}\right).
\end{split}
\end{equation}
We show the base filter in Fourier and configuration space in Figure \ref{fig:morlet}. 

The real components of the Morlet filters are symmetric under a rotation by $\pi$, and the imaginary components are antisymmetric. This means at the complex magnitude step, the resulting scattering fields are also symmetric under a rotation by $\pi$, and therefore only rotations up to an angle of $\pi$ are required. 

\subsection{NFN Formalism}\label{sec:appendix_nfn}
We describe the transformations through the NFN using Einstein summation notation. To produce the complete filter set, we first transform the Fourier space position coordinate $\vec{k}=k^m$ to a set of coordinates corresponding the the scale $j$ and angle $l$ of the filter. This involves first applying the rotation operator $R^{l}_{mn}$, which applies the rotation matrix for a rotation angle of $\theta_l=2\pi l / L$. Then, we apply the dilation operator which we define as $D^{j}=2^{-j}$, corresponding to a dyadic scaling of the coordinates in Fourier space. The resulting transformation is
\begin{equation}
    k^{jl}_{n} =D^{j} R^l_{mn} k^{m}.
\end{equation}
To capture the differences between different scales, we include the scale index $j$ as an additional parameter for the NFN, which allows it to produce different filters for the different scales. We choose this approach instead of defining a new neural field for each scale to allow the NFN to smoothly share information between scales and reduce the necessary parameters. We concatenate $j$ directly into $k^{jl}_{n}$, 
\begin{equation}
    y^{jl}_{n} = \left[k^{jl}_{n}, j\right].
\end{equation} 
We process these coordinates with the NFN to generate $\hat{\psi}^{jl}$, the final filter set in Fourier space. The transformation through the neural field involves multiple layers:
\begin{align}
    h_o^{jl} &= W^{n}_{o} y^{jl}_n + b_o, \\
    h_p^{jl} &= W^{o}_{p} a(h_o^{jl}) + b_p, \\
    \hat{\psi}^{jl} &= W^{p} h_p^{jl} + b.
\end{align}
Here, $h$ represents the output of each hidden layer, with each layer characterized by a weight matrix $W$ and a bias $b$, both of which contain training parameters. These parameters are different for each layer, but notation is excluded to improve readability. The function $a$ introduces nonlinearity into the network, allowing for more complex nonlinear neural fields. We find that using a Rectified Linear Unit (ReLU) activation function consistently results in a collapse to a zero-only output of the NFN, which causes a gradient explosion as the standard deviation across the batch ($\sigma_B$ in equation \ref{eq:loss}) goes to zero. Instead, we use a Leaky ReLU activation function, which converged faster than other smoother activation functions for comparable performance. This two-layer setup is used for all NFNs in this work. The NFN width is the number of neurons in each layer, and is given by the size of the $o$ and $p$ indices. In our tests, the NFN width is allowed to vary, but the size of $o$ always matches the size of $p$. 

\subsection{Verification of the NFST Parity Violation Detection}\label{sec:appendix_verif}
Quantifying the statistical significance of the parity violation is crucial when claiming a detection using the NFST. Broadly, we need to ensure that our parity violation statistic $\mu$ is nonzero due to a true underlying parity violation, rather than random statistical fluctuations in the field.

In practice, we test the null hypothesis that $\mu$ is consistent with zero, given the natural variation that we see due to observing a finite survey, known as the cosmic variance.\footnote{Here, a `survey' is the full test set of patches.} However, in a practical implementation, we are limited to a single survey. We therefore cannot characterize the cosmic variance of $\mu$ by measuring its variability across many different survey volumes. Instead, we bootstrap the computation of $\mu$ and thereby measure its variability across a single survey, and use this as a proxy for the true cosmic variance. In doing this, we make the assumption that the bootstrapped distribution $\mu^*$ is a good approximation of the distribution due to cosmic variance, $\mu^\text{CV}$. 

\begin{figure}
    \centering
    \includegraphics[width=\columnwidth]{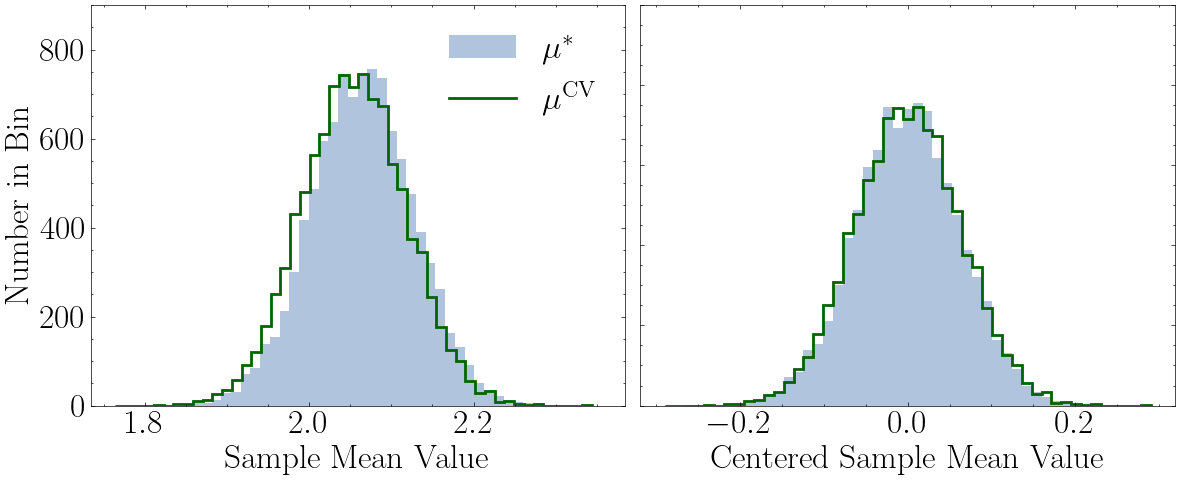}
    \caption{The distributions of the mean of the parity violation detection model output $f(x)$ across a test dataset of 1000 parity violating patches. The bootstrapped distribution $\mu^*$ calculates the mean of $f(x)$ using bootstrapping resampling 10000 times over a single test dataset. The cosmic variance distribution $\mu^\text{CV}$ calculates the mean of $f(x)$ of 10000 independent fully independent test sets. The left panel shows the values directly, showing an expected mismatch between the two distributions. The bootstrapped distribution is computed on one test dataset realization, and should have a mean centred around one of the points in the $\mu^\text{CV}$ distribution. However, the right panel shows the two distributions shifted to mean-zero, where they match well. The variability in $\mu^\text{CV}$ is therefore well estimated by $\mu^*$ and verifies that the bootstrapping is an effective way to estimate cosmic variance of the parity violation detection model.}
    \label{fig:cv_test}
\end{figure}

We test this assumption by generating 10000 full surveys of 1000 patches and building up the distribution $\mu^\text{CV}$ by measuring the value of $\mu$ on each. We then compare this to the bootstrapped distribution $\mu^*$, measured over a single survey of 1000 patches. For this test, we use the best NFST model with an NFN width of 256 for a training and validation size of 400 patches. We show these two distributions in the left panel of Figure \ref{fig:cv_test}. The two distributions do not have the same mean, as we expect because the single survey case will be centred around one of the points in the full cosmic variance distribution. However, if we shift both distributions to zero as shown in the right panel, we find that they are indeed equivalent distributions, in terms of variability. To verify their equivalence, we perform a two-sample KS test using the scipy implementation \texttt{scipy.stats.ks\_2samp}, which tests for the null hypothesis that the two distributions are identical. We find a $p$-value of 0.61, meaning that we cannot reject the null hypothesis, and therefore the distributions are statistically equivalent. The bootstrapping therefore correctly estimates the cosmic variance of $\mu$ for the NFST statistic, for parity violating data. 

\begin{figure}
    \centering
    \includegraphics[width=\columnwidth]{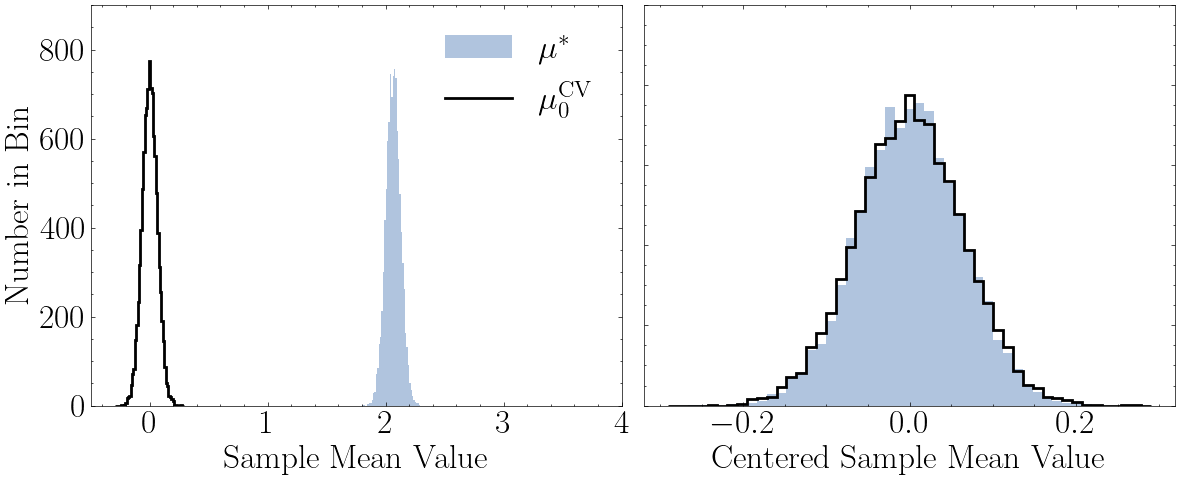}
    \caption{As with Figure \ref{fig:cv_test}, but comparing the bootstrapped distribution $\mu^*$ to the null distribution $\mu_0^\text{CV}$, which calculates the mean of $f(x)$ over 10000 fully independent test sets with no parity violation. The left panel shows that this distribution is centred about 0, because on average the full test sets will have no parity violation signal. The right panel shows the two distributions centered with mean-zero, and shows the variability in the null distribution is well estimated by the bootstrapped distribution. This condition must hold to reject the null hypothesis with bootstrapping alone.}
    \label{fig:ncv_test}
\end{figure}

In this approach, we look to reject the null hypothesis that there is no parity violation by showing that $\mu$ is distinct from zero. We must therefore also ensure that the statistic has the same variability about $\mu=0$ for the case when there is no parity violation in the data. To test this, we construct a parity conserving dataset by placing 8 left-handed and 8 right-handed triangles, maintaining the same total of 16 triangles but without a parity violation. We again measure the distribution of $\mu$ over 10000 full surveys of 1000 patches, and compare it to the bootstrapped distribution from a single parity violating survey. In the left panel of Figure \ref{fig:ncv_test}, we see the full cosmic variance distribution is centred on zero as is expected from data without a parity violation. In the right panel, we again centre both distributions on zero to show that the variability matches. We repeat the KS test, and find a $p$-value of 0.99, again verifying that the distributions are the same. We can therefore use the bootstrapped variability as a measure to reject the null hypothesis that the field is not parity violating.   

Finally, the score $\eta$ is defined as the number of standard deviations of the bootstrapped distribution from zero. For this to be equivalent to a $\sigma$-level of confidence, we have assumed that the distributions are Gaussian. We test for Gaussianity by normalizing each distribution to mean of zero and standard deviation of one, and using the normality test implemented in scipy, \texttt{scipy.stats.normaltest}. In this test, the null hypothesis is that the data follows a normal distribution.  For each of the bootstrapped distribution, the parity violating cosmic variance distribution, and the parity conserving cosmic variance distribution, we find $p$-values of 0.40, 0.84 and 0.53 respectively, showing no evidence against normality. With this result, we verify that $\eta$ is equivalent to a $\sigma$-level of confidence for a parity violation.

Interpreting equation \ref{eq:eta} as similar to a $\chi^2$ test, we can also interpret Figures \ref{fig:cv_test} and \ref{fig:ncv_test} as a verification that the variance is independent of $\mu$, and a verification that the statistic is sufficiently stable that bootstrapping over the data alone is enough to correctly estimate the variance.

Another caveat to this measurement is the choice to take the maximum score after 10 runs. While this allows a fairer comparison by avoiding the case where a model fails to converge, which is a particular issue in the low training data cases, it does mean that $\eta$ may be overestimated. By performing 10 iterations, we improve the chance that the model will perform well on the test set, and hence the interpretation on sigma. For this work however, the core focus is a comparison between models, and since the 10 repeats were consistent across models, $\eta$ still acts as a fair comparison metric. For future work, we will consider more closely the effect that repeating the training process has on the statistical significance of detection.

\end{document}